\RequirePackage{fix-cm}
\documentclass[smallextended]{svjour3}       
\smartqed  
\usepackage{graphicx}
\usepackage[dvipsnames,table,xcdraw]{xcolor}
\usepackage[utf8]{inputenc} 
\usepackage{amsmath,amssymb,authblk}
\usepackage{url,comment}
\usepackage{siunitx}
\usepackage{booktabs}
\usepackage{mathtools}
\usepackage{mhchem}
\usepackage{cancel}
\usepackage{filecontents}
\usepackage{soul}
\usepackage{enumitem}
\usepackage[normalem]{ulem}
\usepackage{tikz}
\usetikzlibrary{external,matrix}
\usepackage[unicode]{hyperref}  
\hypersetup{pdftitle=Multiscale thermodynamics of charged mixtures}
\hypersetup{pdfauthor=}

\newcommand{\MM}{\mathbf{M}}

\newcommand{\uMM}{{\,{}^{\mathbf{M}}}}
\newcommand{\uu}{\mathbf{u}}
\newcommand{\vv}{\mathbf{v}}
\newcommand{\VV}{\mathbf{V}}
\newcommand{\ppi}{\boldsymbol{\pi}}
\newcommand{\umu}{{\,{}^{\ppi}{}}}
\newcommand{\mm}{\mathbf{m}}
\newcommand{\EE}{\mathbf{E}}
\newcommand{\BB}{\mathbf{B}}

\newcommand{\HH}{\mathbf{H}}
\newcommand{\DD}{\mathbf{D}}
\newcommand{\YY}{\mathbf{Y}}
\newcommand{\cDD}{\mathcal{D}}
\newcommand{\bcDD}{\boldsymbol{\mathcal{D}}}
\newcommand{\PP}{\mathbf{P}}
\newcommand{\pp}{\mathbf{p}}
\newcommand{\rF}{\mathrm{F}}
\newcommand{\rr}{\mathbf{r}}

\newcommand{\RR}{\mathbf{R}}
\newcommand{\dd}{\mathrm{d}}
\renewcommand{\AA}{\mathbf{A}}
\newcommand{\xx}{\mathbf{x}}
\newcommand{\yy}{\boldsymbol{\xi}}

\newcommand{\eps}{\varepsilon}

\newcommand{\Lie}{\mathcal{L}}
\newcommand{\oomega}{\boldsymbol{\omega}}
\newcommand{\eepsilon}{\boldsymbol{\epsilon}}
\newcommand{\mmu}{\boldsymbol{\mu}}
\newcommand{\wdagger}{{\widetilde{\dagger}}}
\newcommand{\LambdaM}{\Lambda_{(\MM)}}

\DeclareMathOperator{\curl}{curl}
\DeclareMathOperator{\dive}{div}

\begin{document}

\title{Multiscale thermodynamics of charged mixtures\thanks{This work was supported by the German Research Foundation, DFG project no.~FU~316/14-1, and by the Czech Science Foundation, project no.~17-15498Y.
This work has been supported by Charles University Research program No. UNCE/SCI/023.}
}


\author{Petr V{\'a}gner\and
        Michal Pavelka\and
        O{\u g}ul Esen
}


\institute{P. V{\'a}gner \at
    Weierstrass Institute, Mohrenstrasse 39, 10 117 Berlin, Germany\\
    Tel.: +49 30 20372-390\\
    Fax.: +49 30 20372-317\\
    \and M. Pavelka \at
    Mathematical Institute, Faculty of Mathematics and Physics, Charles University in Prague, Sokolovsk\'{a} 83, 186 75 Prague, Czech Republic\\
    \and O. Esen \at Gebze Technical University, 41400, Gebze-Kocaeli, Turkey
}

\maketitle
\begin{abstract}
 A multiscale theory of interacting continuum mechanics and thermodynamics of mixtures of fluids, electrodynamics, polarization and magnetization is proposed.
The mechanical (reversible) part of the theory is constructed in a purely geometric way by means of semidirect products. This leads to a complex Hamiltonian system with a new Poisson bracket, which can be used in principle with any energy functional.
The thermodynamic (irreversible) part is added as gradient dynamics, generated by derivatives of a dissipation potential, which makes the theory part of the GENERIC framework. 
Subsequently, Dynamic MaxEnt reductions are carried out, which lead to reduced GENERIC models for smaller sets of state variables. Eventually, standard engineering models are recovered as the low-level limits of the detailed theory. The theory is then compared to recent literature.

\keywords{GENERIC \and Electrodynamics \and Continuum mechanics \and Non-equilibrium thermodynamics \and Polarization \and Magnetization \and Multiscale modeling \and Hamiltonian mechanics}
\PACS{05.70.Ln \and 03.50.De}
\subclass{78A25 \and 35Q61 \and 82B35 }
\end{abstract}

%
%
%
\section{Introduction}
Theoretical electrochemistry aims to describe and predict behavior of chemically reacting systems of charged substances.
The modeling methods vary according to the characteristic times, lengths and details of the observed electrochemical systems.
This paper aims to develop a hierarchy of continuum models on different levels of description using the framework of the General Equation for Non-Equilibrium Reversible-Irreversible Coupling (GENERIC) \cite{Grmela1997,ottinger1997dynamics,ottinger2005minimal,pavelka2018multiscale}. 

Let us first briefly recall GENERIC. Consider an isolated system described by state variables $\xx$. The state variables can be for instance position and momentum of a particle, field of probability density on phase space, fields of density and momentum density, electromagnetic fields, etc. Evolution of functionals $F(\xx)$ of the state variables is then expressed as
\begin{align}\label{eq.generic}
	\dot{F}(\xx) = \Big\{ F,E \Big\}%
    + \left\langle \frac{\delta F}{\delta \xx}, \frac{\delta \Xi}{\delta \xx^*} \middle|_{\xx^*=\frac{\delta S}{\delta \xx}}\right\rangle,
\end{align}
where the former term on the right hand side stands for a Poisson bracket of the functional $F$ and the energy $E$ while the latter for scalar product of gradient of $F$ and gradient of the dissipation potential $\Xi$. Conjugate variables (derivatives of the entropy $S$ in the entropic representation) are denoted by $\xx^*$. The Poisson bracket is antisymmetric, which leads to automatic energy conservation, and satisfies Jacobi identity, which expresses intrinsic compatibility of the reversible evolution. The irreversible term yields a generalized gradient flow driven by gradient of entropy and ensures the second law of thermodynamics. Many successful models in non-equilibrium thermodynamics have been formulated in the GENERIC structure \eqref{eq.generic}, and many new thermodynamically consistent models have been obtained by seeking that structure, see e.g. \cite{ottinger2005minimal,pavelka2018multiscale}.

We shall start the present work by recalling the semi-direct product coupling of reversible fluid mechanics and electrodynamics in vacuum \cite{marsden1982hamiltonian}.
In other words, we let the electrodynamics be advected by fluid mechanics as in \cite{pavelka2018multiscale}. To go beyond, we shall add also the field of polarization density and a canonically coupled momentum of polarization. It is important to express the behavior of dipole moment of molecules in interaction with electromagnetic field and the overall motion. To this picture, we shall add magnetization (the famous Landau \& Lifshitz model) advected by fluid mechanics. This way we shall build a hierarchy of levels of description with appropriate Poisson brackets expressing kinematics on the levels, see Figure~\ref{fig:hierarchy}.

Subsequently, we shall introduce dissipation on the most detailed levels of description, we shall present reductions to less detailed (lower) levels down to the level of mechanical equilibrium. On this level the evolution is governed by the
generalized Poisson-Nernst-Planck equations.
We believe that such a complete and geometric picture of continuum thermodynamics of matter coupled with electrodynamics (including polarization and magnetization) was missing in the literature.

\subsection{Mathematical  aspects, terminology and further remarks}
A description in terms of fields describing physical quantities, e.g. mass density, magnetic field, is developed throughout the manuscript. A mathematically rigorous formulation of the Poisson brackets and the dissipation potentials for the fields would require deployment of an advanced functional analysis, see e.g.~\cite{marsden2013introduction}. We opt-out from discussing the setting of the function spaces, convergence of the integrals and other mathematical aspects that are not pertinent to the discussed physics at this stage of the development. 

Only the bulk equations are investigated in this manuscript, and all boundary terms are thus neglected in the calculations below.
This means that the considered systems are assumed to be either boundary-less, isolated or (infinitely) large with a reasonable decay of the fields at infinity.
This is a standard assumption in geometric field theories, see e.g.~\cite{ottinger2005minimal,pavelka2018multiscale,marsden2013introduction}.

A level of description is given by a set of variables. If two sets of variables, belonging to the respective levels of description, can be ordered by inclusion, then we say that the level of description with smaller set of variables is less detailed level of description and vice versa. 

This brings us to an another interpretation of the multiscale description not necessarily connected to different space-time scales. For instance in the Grad hierarchy~\cite{grad,miroslav-grad}, where kinetic moments are considered \cite{struch}, the scale (or rather level of description) is given by the number of moments considered. The advantage of the latter meaning is that it is independent of the particular non-objective and observer-dependent spatio-temporal scale.

The term~\textit{reversible} is used through out the article in connection with the structure of the evolution equations in the sense of the time-reversal transformation~\cite{pavelka2014time}. The Hamiltonian systems may often be cast in to the first-order, symmetric hyperbolic partial differential equation form~\cite{peshkov2018continuum}. Assuming smooth initial data, solutions of the hyperbolic systems may be constructed for short times using the Cauchy-Kowalevskaya theorem~\cite{evansPDE}. Such solutions are then short-time reversible in the usual sense of partial differential equations. 

\subsection{Notation $(\partial_t\xx)_\text{rev/irr}$ and $\xx^\dagger$, $\xx^*$}\label{sec:notation}
A problem under consideration usually admits multiple equivalent choices of state variables. Let us denote the family of the admissible, equivalent, levels of description as $\mathbb{X}$.
For a given level of description, characterized by state variables $\xx\in\mathbb{X}$, the presented theory aims to describe evolution of functionals depending on $\xx$, i.e.\ we are interested in $\dot F(\xx)$, see~\eqref{eq.generic}.
Therefore, functional derivatives
\begin{align}
    \left(\frac{\delta F}{\delta \xx}\right)  \equiv F_\xx~
\end{align}
are often used in this manuscript.

The part of the evolution generated by the Poisson bracket $\{\cdot,\cdot\}$ and the energy $E(\xx)$, see~\eqref{eq.generic}, will be called \textit{Hamiltonian} or \textit{reversible} and denoted as
\begin{align}
 \left\{F,E\right\} = \left( \dot F \right)_\text{rev} = \left(\partial_t F\right)_\text{rev}~.
\end{align}
The functional derivative of the energy w.r.t.\ variable $\xx$, also called the \textit{energy-conjugate} to the variable $\xx$, will be briefly denoted by $\dagger$ in the superscript, i.e.\ $E_{\xx} = \xx^\dagger$. 
Poisson brackets are usually conveniently expressed in the \textit{energetic representation}, i.e.\ when entropy density $s$ is amongst the state variables, $(s, \yy)\in\mathbb{X}$, where $\yy$ denotes the state variables other than the entropy density.

The part of the evolution generated by the dissipation potential $\Xi$ and the entropy $S(\xx)$, see~\eqref{eq.generic}, will be called \textit{irreversible} and denoted as
\begin{align}
\left\langle F_\xx, \Xi_{\xx^*}|_{\xx^*=S_\xx}\right\rangle  = \left( \dot F \right)_\text{irr} = \left(\partial_t F\right)_\text{irr}~.
\end{align}
The symbol $\xx^*$, called \textit{entropy-conjugate}, denotes the variables on which the dissipation potential $\Xi(\xx^*)$ depends. 
The notation $\xx^*$ is going to be overloaded in Section~\ref{sec:dissipation} because it will be also used to denote the value of the entropy derivative, i.e.\ $\xx^* = S_\xx$.
Since the irreversible evolution is expressed in terms of the derivatives of the dissipation potential w.r.t.\ $\xx^*$, presence of energy density $e$ amongst the variables makes it simple to control the energy conservation, i.e., $(e,\yy)\in\mathbb{X}$. Such set of variables is called \textit{entropic representation} \cite{callen}.

The relations between the energy-conjugates in the energetic representation $(s^\dagger, \yy^\dagger)$ and the entropy-conjugate in the entropic representation $(e^*, \yy^*)$ are explained in \cite{pavelka2018multiscale,CMAT2018}.

The overall evolution of a functional $F$ is then composed from the reversible and irreversible parts, 
\begin{align}
    \partial_t F = \left(\partial_t F\right)_\text{rev} + \left(\partial_t F\right)_\text{irr}~.
\end{align}
Symbols $(\partial_t F)_\text{rev/irr}$ be interpreted as the reversible and irreversible (w.r.t\ TRT) parts of the right hand side of the evolution equations, or as the reversible and irreversible vector fields, see e.g.~\cite{pavelka2014time}.

\section{Hierarchy of Poisson brackets}
\subsection{Hamiltonian dynamics and semidirect product theory}

A physical system described by state variables $\mathbf{x}$ is said to be Hamiltonian if the differential equations governing its dynamics can be written in the form of
\begin{align} \label{HamEq}
    (\dot{\mathbf{x}})_\text{rev}=\{\mathbf{x},E\},
\end{align}
where dot stands for (typically partial) time derivative, and $E$ is a Hamiltonian (energy)
function \cite{marsden2013introduction,de2011methods,Marsden1983}. Here, the (Poisson) bracket
\begin{align*}
\{\bullet,\bullet \}:\mathcal{F}\times \mathcal{F} \longrightarrow \mathcal{F}
\end{align*}
is a (skew-symmetric and bilinear)
mapping on the space of smooth functions depending on the state variables $\mathbf{x}$. It is required that a Poisson bracket must satisfy both the Leibniz and the Jacobi identities \cite{weinstein1998poisson} that is
\begin{subequations}
\begin{align}
&\{FG,H\}=\{F,H\}G+F\{G,H\}, \\
&\{\{F,G\},H\}+\{\{G,H\},F\}+\{\{H,F\},G\}=0,
\end{align}
\end{subequations}
for any real valued functions $F$, $G$ and $H$. 
A space endowed with such a bracket is called a Poisson space \cite{abraham1978foundations,Marsden-Ratiu-book,libermann2012symplectic}.
 Skew-symmetry of the bracket
manifests conservation of the Hamiltonian function all along the motion. This structure implies in particular to conservation of energy. This observation is fundamental for reversible character of the Hamiltonian systems. More on time-reversibility and Onsager-Casimir reciprocal relations can be found in \cite{pavelka2018multiscale}.
\bigskip

\noindent In order to write a system of equations in the
Hamiltonian form \eqref{HamEq}, two tasks must
simultaneously be accomplished after deciding the state variables \cite{abraham1978foundations}. These are (1) to choose a Hamiltonian function and (2) to construct a Poisson bracket. Even though there is no algorithmic way to accomplish these tasks, there are some  techniques to perform Hamiltonian analysis of a given system. Let us depict here one of them that we use in the present work. Start with the most basic form of the system (by removing all possible field
extensions and symmetries). After the Hamiltonian analysis of the basic model is established, field extensions can be added by applying pure geometric and algebraic techniques. For example, consider a classical hydrodynamical system, start initially with analysis of  incompressible and isentropic fluid flow. After this is done, let the mass density and the entropy vary in time. Additions of these scalar fields can be done in purely abstract framework by using the semi-direct product theory \cite{marsden-ratiu-weinstein,esen2017hamiltonian}. Accordingly, in this work, we shall start with Hamiltonian realizations of two basic models, namely the one-component compressible fluid mechanics. Then we extend these models with the additions of some field theories such as electromagnetic fields, polarization and magnetization. 

\bigskip

\noindent For the basic models (without field extensions and symmetries), in the literature, 
there exist some popular Poisson structures to begin with. Let us comment on these
Poisson structures one by one in order to fix the notation and to make the present work more complete. 

\paragraph{Canonical Poisson bracket.} 
Consider a vector space $V$ and its linear algebraic dual $V^*$. In finite dimensions, the duality (pairing) $\langle \bullet, \bullet \rangle$ between $V$ and $V^*$ can be considered as the Euclidean scalar (dot) product whereas, in infinite dimensions, (that is if $V$ is a function space or the space of some fields) the duality can be an $L^2$-pairing (that is simply multiply-and-integrate form). The cotangent bundle $T^{\ast }V$ is isomorphic to the product space $V\times V^{\ast }$ and carries a canonical Poisson structure. Let us exhibit this geometry. The elements of the cotangent bundle $T^*V$ are pairs $(a,a^*)$ in $T^{\ast }V$. Consider two function(al)s $F$ and $H$ on $T^*V$, that depend on $(a,a^*)$. By referring to the duality between $V$ and $V^*$, the canonical Poisson bracket is defined to be
\begin{align} \label{cpb}
\{F,H\}^{T^*V}(a,a^*)=\langle F_a, H_{a^*} \rangle - \langle H_a, F_{a^*} \rangle, 
\end{align}
where, for example, $F_a$ denotes the partial (for infinite cases, Fr\'{e}chet) derivative of $F$ with respect to $a$. We assume that the vector space is reflexive that is $V^{**}=V$. This enables us consider $F_a$ as an element of $V^*$ for $a$ in $V$, and $F_{a^*}$ as an element of $V$ for $a^*$ in $V^*$.    

\paragraph{Lie-Poisson bracket.}
Let us start once more with a vector space $\mathfrak{g}$. In this case, we are not coupling $\mathfrak{g}$ with its dual as in the previous case, but we consider $\mathfrak{g}$ as a Lie algebra. A vector space $\mathfrak{g}$ is called a Lie algebra if it admits a bilinear skew-symmetric mapping (Lie bracket) 
\begin{align}
[\bullet,\bullet]: \mathfrak{g}\times \mathfrak{g}\longrightarrow  \mathfrak{g}
\end{align}
satisfying the Jacobi identity. In this case, linear algebraic dual $\mathfrak{g}^*$ of $\mathfrak{g}$ carries a Poisson bracket called as the Lie-Poisson bracket. More concretely, if $F$ and $H$ are two function(al)s on $\mathfrak{g}^*$ then, for $\mu$ in $\mathfrak{g}^*$,
\begin{align} \label{LPb}
\{F,H \}^{\mathfrak{g}^*}(\mu)=\pm\langle \mu, [F_\mu, H_\mu] \rangle
\end{align}
where $F_\mu$ is assumed to be an element of $\mathfrak{g}$ whereas $\langle \bullet, \bullet \rangle$ is the pairing between $\mathfrak{g}^*$ and $\mathfrak{g}$. Notice that there exist two Lie-Poisson brackets according to the sign of the bracket (\ref{LPb}). The choice of the sign is not arbitrary for a chosen physical system. It is minus for the rigid body dynamics, since it admits a symmetry due to the left action that is, matrix multiplication of the rotation group $SO(3)$ on the configuration space $\mathbb{R}^3$, see \cite{Fecko,Marsden-Ratiu-book}. In this present work, we employ the plus Lie-Poisson bracket since, in the continuum theories there exists a right symmetry, physically called as the particle relabeling symmetry \cite{MaWe81,Marsden1983}. We will continue to keep plus minus notation in this introductory subsection.  
After having presented the canonical and Lie-Poisson structures, let us now briefly explain (both direct and semi-direct) couplings and extensions of these geometries.

\paragraph{Direct product of two Poisson structures.} 
The most direct way to couple two Poisson structures is to add the associated Poisson brackets directly without any cross terms and interactions. Consider for example two Poisson spaces $P_1$ and $P_2$ equipped with Poisson brackets $\{\bullet,\bullet \}^{P_1}$ and $\{\bullet,\bullet \}^{P_2}$, respectively. Then the direct product Poisson structure is 
\begin{equation} \label{dp-PB}
\{F,H\}^{P_1\times P_2}=\{F,H\}^{P_1}+\{F,H\}^{P_2}
\end{equation}
for two function(al)s on $P_1\times P_2$. In particular, consider two Lie
algebras $\mathfrak{g}_1$ and $\mathfrak{g}_2$, and the dual spaces $\mathfrak{g}^*_1$ and $\mathfrak{g}^*_2$, respectively. Then by employing the isomorphism $(\mathfrak{g}_1\times\mathfrak{g}_2)^*=\mathfrak{g}_1^*\times\mathfrak{g}_2^*$, we construct the Lie-Poisson bracket on the (direct product) dual space by 
\begin{align} \label{LPb-dp}
\{F,H \}^{\mathfrak{g}_1^*\times \mathfrak{g}_2^*}=\{F,H \}^{\mathfrak{g}^*_1}+\{F,H \}^{\mathfrak{g}^*_2},
\end{align}
where the Lie-Poisson brackets on the right hand side are the ones on $\mathfrak{g}_1^*$ and $\mathfrak{g}_2^*$, respectively. 
   

\paragraph{Semi-direct product extension.} 
We now extend the Lie-Poisson bracket. Consider a Lie algebra $\mathfrak{g}$ and a vector space $V$. Assume that $\mathfrak{g}$ acts on $V$, that is there exist bilinear mapping
\begin{align} \label{act}
\rho:\mathfrak{g} \times V\mapsto V.
\end{align}  
By fixing an element $\xi$ in $\mathfrak{g}$, we define a linear mapping $\rho_\xi$ on $V$ \cite{Hall-LA-book}. This enables us to define a Lie algebra bracket on the product space $\mathfrak{g}\times V$ given by
\begin{align}
\left[ \left( \xi _ { 1 } , v _ { 1 } \right) , \left( \xi _ { 2 } , v _ { 2 } \right) \right] = \left( \left[ \xi _ { 1 } , \xi _ { 2 } \right] , \rho_{\xi _ { 1 }} (v _ { 2 }) - \rho_{\xi _ { 2 }} (v _ { 1 }) \right).
\end{align}
By referring to the Lie-Poisson structure presented in the previous paragraph, the dual space $\mathfrak{g}^*\times V^*$ turns out to be a Poisson space equipped the Lie-Poisson bracket. For two function(al)s $F$ and $H$ depending on $(\mu,a^*)$ in $\mathfrak{g}^*\times V^*$, the Lie-Poisson bracket is computed to be   
\begin{equation} \label{sdp-1}
\{ F , H \}^{\mathfrak{g}^*\times V^*}( \mu , a^* ) = \pm \left\langle \mu , \left[  F_\mu  , H_\mu \right] \right\rangle \pm \left\langle {a^*} , \rho_{F_\mu}(H_{a^*})- \rho_{H_\mu}(F_{a^*}) \right\rangle,
\end{equation}
where the first pairing on the right hand side is the one between $\mathfrak{g}^*$ and $\mathfrak{g}$ whereas the second pairing on the right hand side is the one between $V^*$ and $V$.

\paragraph{Direct product coupling of the canonical Poisson bracket and the Lie-Poisson bracket.} 
Consider two Poisson structures, the canonical Poisson bracket \eqref{cpb} on $T^*V$ and the Lie-Poisson bracket \eqref{LPb} on $\mathfrak{g}^*$. If a physical system has a phase space as the product of these two spaces $\mathfrak{g}^*\times T^*V$, then one way to define a Poisson bracket on the product space by simply adding two brackets \eqref{cpb} and \eqref{LPb}. This is, for two function(al)s $F$ and $H$ depending on $(a, a^*,\mu)$, given by
\begin{align} \label{direct}
\{F,H\}^{T^*V\times \mathfrak{g}^*}_{(DP)}= \{F,H\}^{T^*V} \pm \{F,H\}^{\mathfrak{g}^*}. 
\end{align}

\paragraph{Semi-direct product coupling of the canonical Poisson bracket and the Lie Poisson bracket.}  If the quantities in $V$ are advected by the quantities in $\mathfrak{g}$, that is if there is an action of $\mathfrak{g}$ as given in \eqref{act}, then the Poisson bracket in \eqref{direct} fails to be the correct Poisson structure for the system. In this case, the cross terms (see, for example, the second term in \eqref{sdp-1}) must be added as in the semi-direct product theory exhibited in the previous paragraph. In this case,  one is equipped, in addition to the action \eqref{act}, with a (dual) action of $\rho^*_\xi$ on the dual space $V^*$. This is due to the preservation of the duality between $V$ and $V^*$ under the action, 
\begin{align}
\langle a^*, \rho_\xi(a)\rangle=\langle \rho^*_\xi(a^*), a\rangle
\end{align}
for all $a$ in $V$ and $a^*$ in $V^*$. Note that the dual action $\rho^*$ is computed as linear algebraic dual of the linear mapping $\rho_\xi$. A direct computation shows that for two function(al)s $F$ and $H$, under the existence of the actions, the Poisson bracket is  
\begin{align} \label{Can-LP-coupling}
    \{F,H\}^{T^*V\times \mathfrak{g}^*}_{(SDP)} ={}& \{F,H\}^{T^*V} \pm \{F,H\}^{\mathfrak{g}^*} \\ &\pm \left\langle a^* , \rho_{F_\mu}(H_{a^*})- \rho_{H_\mu}(F_{a^*}) \right\rangle \pm \left\langle a , \rho_{F_\mu}^*(H_a)- \rho_{H_\mu}^*(F_a) \right\rangle.\nonumber
\end{align}
Notice that the first line is the same with the direct product case (non interaction) in \eqref{direct}, and the second line is a manifestations of the advection. This is a particular case of the matched pair of Hamiltonian dynamics presented in \cite{ogul2016matched2}.
   
\subsubsection{Matched pairs}
Only one sided couplings and semidirect products are considered in this work. Only recently the two-sided actions have been developed in classical physics, called Matched pairs~\cite{ogul2016matched2}. In the Galilean setting the one-sided actions are usually enough, but it is indeed possible that matched pairs would turn necessary for the fully relativistic and quantum (see works by Majid~\cite{Majid-book}) treatment. So far the only physically important matched pair in classical physics has been identified in Grad hierarchy, where coupling of fluid mechanics and higher kinetic moments plays an important role \cite{momentum-Euler}.

\subsection{Hamiltonian fluid mechanics of mixtures}
The Poisson bracket expressing kinematics of fluid mechanics has been long known \cite{arnold,dv,GrPD,MG84,Morrison-brackets}. The Poisson bracket can be easily extended to mixtures with multiple densities, momenta and entropies (i.e. temperatures), see e.g. \cite{hierarchy}.

\subsubsection{Classical fluid mechanics}\label{sec.FM}
The state variables of classical fluid mechanics are fields of density, momentum density and entropy density, $\xx = (\rho,\uu,s)$. The Poisson bracket generating one-component compressible fluid mechanics (hydrodynamic Poisson bracket) is
\begin{align}\label{FM01}
\{F,G\}^{(\text{FM})}(\rho, \uu, s)%
&= \int \dd\rr \rho \left(\partial_i F_\rho G_{u_i}-\partial_i G_\rho F_{u_i}\right)\nonumber\\
&+ \int \dd\rr u_i \left(\partial_j F_{u_i} G_{u_j}-\partial_j G_{u_i} F_{u_j}\right)\nonumber\\
&+ \int \dd\rr s \left(\partial_i F_s G_{u_i}-\partial_i G_s F_{u_i}\right),
\end{align}
where $\rho$, $\uu$ and $s$ are \textit{mass density}, \textit{mass momentum density} and \textit{volumetric entropy density}, respectively. The Poisson bracket in (\ref{FM01}) is also an example of a semidirect product bracket obeying the form exhibited in (\ref{sdp-1}). This well-know analysis can be found, for example, in \cite{arnold,marsden-ratiu-weinstein,be}. 

For an arbitrary energy $E(\rho, \uu, s)$, the reversible evolution of a functional $F$ of the state variables reads
\begin{align}
    (\dot{F})_\text{rev}=&\{F,E\}^{(\text{FM})}\nonumber\\
	=& \int\dd\rr F_{\rho}\left(-\partial_i( \rho u^{\dagger i})\right)\nonumber\\
	& +\int\dd\rr F_{u_i}\left(-\rho\partial_i \rho^\dagger - u_j \partial_i u^{\dagger j} - s\partial_i s^\dagger - \partial_j(u_i u^{\dagger j})\right)\nonumber\\
	& +\int\dd\rr F_{s}\left(-\partial_i( s u^{\dagger i})\right),
\end{align}
where integration by parts was used several times\footnote{The energy-conjugate to variable $\xx$, i.e. functional derivative of the energy E w.r.t. $\xx$, is going to be denoted by $\xx^\dagger$, i.e.\ $E_\xx = \xx^\dagger$. The notation is summarized in section~\ref{sec:notation}.}.
Boundary terms disappear as we assume isolated (e.g. periodic) system. 

By comparing with the chain rule
\begin{align}
	\dot{F} = \int\dd\rr\left(F_\rho \partial_t \rho + F_{u_i}\partial_t u_i +F_s \partial_t s\right)
\end{align}
we can read the reversible evolution equations for fluid mechanics,
\begin{subequations}
\begin{align}
    (\partial_t \rho)_\text{rev} =& -\partial_i( \rho u^{\dagger i})~,\\
    (\partial_t u_i)_\text{rev} =& -\rho\partial_i \rho^\dagger - u_j \partial_i u^{\dagger j} - s\partial_i s^\dagger - \partial_j(u_i u^{\dagger j})~,\\
    (\partial_t s)_\text{rev}  =& -\partial_i( s u^{\dagger i})~.
\end{align}
\label{FM03}
\end{subequations}
For the usual choice of energy, 
\begin{align}
    E^{\text{Euler}}\left(\rho,\uu,s \right) = \int\dd\rr \left(\frac{\uu^2}{2\rho}+\eps(\rho,s)\right),
  \label{FM02}
\end{align}
the compressible non-isothermal Euler equations for ideal fluids are obtained from the system~\eqref{FM03}.
The energy conjugates $\rho^\dagger$, $\uu^\dagger$ and $s^\dagger$ represent the chemical potential, barycentric velocity and temperature, respectively.

\subsubsection{Hamiltonian fluid mechanics of mixtures}
Consider now a mixture of $n+1$ species, each of which is described by its own density, momentum density and entropy density. 
The Poisson bracket expressing kinematics of state variables $\xx=(\rho_\alpha,\uu^\alpha,s_\alpha)$, $\alpha \in \{0,1,\cdots,n\}$, is
\begin{align} 
    \label{MIX01}%
    \{F, G\}(\rho_\beta, \uu^\beta, s_\beta) = \sum_{\alpha=0}^n \{F,G\}^{\textrm{(FM)}_\alpha}(\rho_\alpha, \uu^\alpha, s_\alpha)~.
\end{align}
This Poisson bracket can be derived for instance by projection from the Liouville equation \cite{hierarchy}.
It consists of the sum of $n+1$ Poisson brackets~\eqref{FM01}, each expressed in terms of variables of mixture component $\alpha$.  The Poisson bracket in (\ref{MIX01}) is a direct product Poisson bracket fitting the abstract framework in (\ref{dp-PB}).

Poisson bracket \eqref{MIX01} depends on $n+1$ momenta and $n+1$ entropies, each for one component of the mixture, which is a rather detailed description allowing for independent motion of the constituents and for different temperatures of them (as in cold plasma, where electrons have different temperature than ions). We are, however, often interested in less detailed description, keeping only densities of the species, the total mass momentum and the total entropy,
\begin{align}
    \uu = \sum_{\alpha=0}^n \uu^\alpha \qquad \mbox{and} \qquad s = \sum_{\alpha=0}^n s_\alpha \label{MIX03}.
\end{align}
By letting the arbitrary functional depend only on state variables $\xx=(\rho_\alpha,\uu,s)$, bracket \eqref{MIX01} becomes 
\begin{align}
    \{F,G\}^{\textrm{(CIT)}} &= 
    \sum_{\alpha=0}^n \int \dd\rr \rho_\alpha \left(\partial_i F_{\rho_\alpha} G_{u_i}-\partial_i G_{\rho_\alpha} F_{u_i}\right)\nonumber\\
&+ \int \dd\rr u_i \left(\partial_j F_{u_i} G_{u_j}-\partial_j G_{u_i} F_{u_j}\right)\nonumber\\
&+ \int \dd\rr s \left(\partial_i F_s G_{u_i}-\partial_i G_s F_{u_i}\right)\label{CIT01},
\end{align}
which is referred to as the classical mixture hydrodynamic bracket. It generates the reversible part of Classical Irreversible Thermodynamics (CIT) \cite{degroot1984nonequilibrium}.

The descriptions of the fluid and fluid mixture dynamic considered in the remainder of the paper will be based on the brackets~\eqref{FM01} and~\eqref{CIT01}, respectively.
Hence, the description considering the distinguished momenta of species~\eqref{MIX01} is further avoided.

\subsection{Electrodynamics in vacuum}\label{sec.elmag}
The reversible evolution of electromagnetic fields is generated by the canonical Poisson bracket, see \cite{marsden1982hamiltonian,holm1986hamiltonian,pavelka2018multiscale},
\begin{align}\label{VED01}
	\{F,G\}^{\textrm{(EM)}_\textbf{A}}(\AA, \YY) = \int \dd\rr \left(F_{A_i} G_{Y^i}-G_{A_i} F_{Y^i}\right)
	= \int \dd\rr \left(F_{D^i} G_{A_i}-G_{D^i} F_{A_i}\right),
\end{align}
where $\textbf{A}$ stands for the \textit{vector potential} and $\YY=-\DD$ denotes negative of the \textit{electric displacement field} (either in variables $(\AA,\YY)$ or $(\AA,\DD)$). Poisson bracket in \eqref{VED01} is the canonical Poisson bracket as in (\ref{cpb}). Here, the vector space $V$ is consisting of vector potentials whereas the dual space $V^*$ is the space of electric displacement fields. 

Let us define the \textit{magnetic field} $\BB$ as
\begin{align}
    B^i = \varepsilon^{ijk} \partial_j A_k.
    \label{EM01}
\end{align}
In order to express the bracket~\eqref{VED01} in terms of magnetic field, we assume that the functionals depend only on the curl of $\AA$.
Bracket~\eqref{VED01}, transformed in terms of $(\DD, \BB)$, see~\cite{esen2017hamiltonian}, becomes 
\begin{align}\label{VED02}
    \{F,G\}^{\textrm{(EM)}}(\DD, \BB) = \int \dd\rr \left(F_{D^i}\varepsilon^{ijk}\partial_j G_{B^k}-G_{D^i}\varepsilon^{ijk}\partial_j F_{B^k}\right).
\end{align}
This is the Poisson bracket expressing kinematics of electromagnetic fields $\DD$ and $\BB$.

For an arbitrary energy $E$ the evolution equations of the electromagnetic field given by~\eqref{VED02} are
\begin{subequations}
\label{EM05}
\begin{align}
    (\partial_t D^i)_\text{rev} = {}& \varepsilon^{ijk} \partial_j B_k^\dagger~,\\
    (\partial_t B^i)_\text{rev} = {}&-\varepsilon^{ijk} \partial_j D_k^\dagger~,
\end{align}
where daggers again denote the corresponding derivatives of energy. 
\end{subequations}
The conjugates are interpreted as electric field and magnetic intensity in the laboratory frame,
\begin{equation}
\EE = \frac{\delta E}{\delta \DD} 
\qquad\mbox{and}\qquad
\HH = \frac{\delta E}{\delta \BB}.
\end{equation}
A concrete energy will be specified later after matter is added to the system so that the theory becomes Galilean invariant.

Applying divergence to \eqref{EM05} gives the following evolution equations:
\begin{align}
\partial_t\dive\BB = 0 \qquad \mbox{and} \qquad \partial_t \dive \DD = 0~.
\label{VED04}
\end{align}
The first equality holds due to~\eqref{EM01}, the second equality demonstrates the absence of charges in a vacuum.
Hence, the usual constraints--Gau{\ss}'s laws ~\cite{jackson1998classical}--hold true if satisfied by the initial condition, see~\cite{marsden1982hamiltonian}. Equations \eqref{VED04} are thus a consequence of evolution equations \eqref{EM05}.

\subsection{Electromagnetic field advected by charged fluids}
The purpose of this section is to formulate coupled kinematics of fluids and electromagnetic fields. We employ the theory of semidirect product to find such coupling, and then we perform a transformation unveiling the usual form of the Lorentz force.

\subsubsection{Semidirect product extension}\label{sec.SP}

Let us recall the Lie-Poisson structure in (\ref{CIT01}) expressing kinematics of Classical Irreversible Thermodynamics, i.e. for state variables $(\rho,\uu,s)$, and the \emph{canonical} Poisson bracket in \eqref{VED02} for the electromagnetic field theory in terms of the fields $(\DD, \BB)$. In this Subsection, we couple these two in the light of the abstract framework in (\ref{Can-LP-coupling}) assuming that the action is minus the Lie derivative $-\Lie$ with respect to the velocity field $\uu^\dagger$. The physical meaning is that a quantity is Lie-dragged or advected by velocity of the continuum (conjugate to the total momentum), see e.g. \cite{arnoldbook,LagEul}. We obtain Poisson bracket
\begin{align}
\label{eMIX05}\left\{F,G\right\}^{\textrm{(mEMHD)}}(\rho_\alpha, \mm, s, \DD,& \BB)=\{F,G\}^{\textrm{(CIT)}}|_{\uu=\mm} + \{F,G\}^\text{(EM)}\\
\{F,G\}^{\textrm{(SP)}}(\DD,\mm)&
\begin{dcases}
  &+\int\dd\rr D^i\left(\partial_j F_{D^i} G_{m_j} - \partial_j G_{D^i}F_{m_j}\right)\nonumber\\
  &+\int\dd\rr \partial_j D^j(F_{m_i} G_{D^i} -G_{m_i} F_{D^i})\nonumber\\
  &+\int\dd\rr D^j(  F_{m_i} \partial_j G_{D^i} - G_{m_i} \partial_j F_{D^i})\nonumber\\
\end{dcases}\\
\{F,G\}^{\textrm{(SP)}}(\BB,\mm)&
\begin{dcases}
&+\int\dd\rr B^i\left(\partial_j F_{B^i} G_{m_j} -\partial_j G_{B^i}F_{m_j}\right)\nonumber\\
&+\int\dd\rr \partial_j B^j(F_{m_i} G_{B^i} -G_{m_i} F_{B^i})\nonumber\\
&+\int\dd\rr B^j(F_{m_i} \partial_j G_{B^i} - G_{m_i} \partial_j F_{B^i}),
\end{dcases}
\end{align}
where $\mm$ denotes total momentum density (of matter and electromagnetic field),
\begin{align}
\label{eMIX04}\mm = \uu + \DD\times\BB~.
\end{align}
This standard relation, see e.g. \cite{Landau2}, can be either verified by a direct calculation~\cite{pavelka2018multiscale} or inferred using physical arguments, see Appendix~\ref{app:momentum}.
Note that the notation introduced in~\eqref{eMIX05} allows to write briefly
\begin{align}
  \left\{F,G\right\}^{\textrm{(mEMHD)}}(\rho_\alpha, \mm, s, \DD,& \BB)=\{F,G\}^{\textrm{(CIT)}}|_{\uu=\mm} + \{F,G\}^\text{(EM)}\nonumber \\
  &+\{F,G\}^{\textrm{(SP)}}(\DD,\mm) + \{F,G\}^{\textrm{(SP)}}(\BB,\mm)
\label{eMIX07}
\end{align}
Bracket \eqref{eMIX05} expresses kinematics of a CIT mixture and electromagnetic field with state variables $\xx=(\rho_\alpha,\mm,s,\DD,\BB)$ and was found (for the single species case) in \cite{ADER,CMAT2018,holm1986hamiltonian}.

\subsubsection{Transformation to mass momentum}\label{sec:mass_momentum}
Let us suppose that each mixture component carries charge $e_0\frac{z_\alpha}{m_\alpha}$ proportional to the free charge density,
 which is defined as
\begin{align}
    \rho_\textrm{f} = \sum_{\alpha=1}^n \frac{z_\alpha e_0}{m_\alpha}\rho_\alpha.
    \label{eMIX00}
\end{align}
Suppose functionals $F(\rho_\alpha, \mm , s, \DD, \BB) = \widetilde F(\rho_\alpha, \uu , s, \DD, \BB)$, the relation of the total and mass momenta~\eqref{eMIX04} implies the following transformation rules between their derivatives, it reads
\begin{subequations}
\begin{align}
    F_{\rho_\alpha} = \widetilde F_{\rho_\alpha},\ F_s = \widetilde F_s,~F_\mm = \widetilde F_\uu~,\\
    F_{D^i} = \widetilde F_{D^i} + \varepsilon_{ijk}\widetilde F_{u_j} B^k~,\label{eMIX15b} \\
    F_{B^i} = \widetilde F_{B^i} - \varepsilon_{ijk}\widetilde F_{u_j} D^k~, 
\end{align}
    \label{eMIX15}
\end{subequations}
see Appendix~\ref{app:momentum} for further details. Poisson bracket \eqref{eMIX05} can be now transformed by means of relations~\eqref{eMIX15} to the mass momentum $\uu$ instead of the total momentum $\mm$. The calculation was carried out in \cite{esen2017hamiltonian} and \cite{pavelka2018multiscale} and leads to Poisson bracket
\begin{align}
  \left\{F,G\right\}^{\textrm{(uEMHD)}}&(\rho_\alpha, \uu, s, \DD, \BB)=\{F,G\}^{\textrm{(CIT)}} + \{F,G\}^{\textrm{(EM)}}\nonumber\\
  {}&+\int\dd\rr \sum_{\alpha=1}^n e_0\frac{z_\alpha\rho_\alpha}{m_\alpha}\left(F_{u_i} G_{D^{i}} -  G_{u_i}F_{D^{i}}\right)\nonumber\\
  {}&+\int\dd\rr \sum_{\alpha=1}^n e_0\frac{z_\alpha\rho_\alpha}{m_\alpha}B^i \varepsilon_{ijk}F_{u_{j}} G_{u_{k}} \label{eMIX03},
\end{align}
which is the Poisson bracket expressing evolution of a CIT mixture coupled with electromagnetic field (using the mass momentum $\uu$).\footnote{the transformation crucially depends on $\dive\BB = 0$}

Energy $E(\rho_\alpha, \mm , s, \DD, \BB, \EE, \HH)$ when considering $\mm$ as a state variable is equal to energy $\widetilde{E}(\rho_\alpha, \uu , s, \DD, \BB)$ when considering $\uu$ as a state variable. However, to distinguish between the conjugates, we introduce the notation $\widetilde{E}_{\xx} = \xx^\wdagger$. It follows from the formulae~\eqref{eMIX15} that the chemical potentials, temperature and velocity are unaffected by transformation~\eqref{eMIX04}, i.e. 
\begin{subequations}
\begin{align}
    \rho_\alpha^\dagger=\rho_\alpha^\wdagger,\
    s^\dagger=s^\wdagger,\quad\text{and}\quad
    \vv = \mm^\dagger =\uu^\wdagger~.
\end{align}
The energy-conjugates to $\DD$ and $\BB$ read
\begin{align}
    \label{eMIX17}
    \DD^\dagger = \DD^\wdagger + \uu^\dagger\times\BB~,\\
    \BB^\dagger = \BB^\wdagger - \uu^\dagger\times\DD~,
\end{align}
\end{subequations}
respectively. Note that the electromagnetic fields on the left hand side are seen from the fluid frame while those on the right hand side from the inertial lab frame, see \ref{sec.Gal} for more details.

Energy can be chosen as 
\begin{equation}\label{eq.E.EMHD.usual}
E_{EMHD} = \int\dd\rr \left(\frac{1}{2}\left(\sum_\alpha \rho_\alpha\right) \vv^2 + \frac{1}{2}\left(\DD\cdot\EE + \BB\cdot\HH\right) + \eps(\rho_a, s)\right),
\end{equation}
c.f. \cite{van-magnetic}.
The first part represents kinetic energy of the center of mass. The second part is the electromagnetic energy, see e.g. \cite{feynman2}, where all the fields are measured in the laboratory frame. Finally, the last part is the internal energy. In order to express the energy in terms of the state variables $(\rho_\alpha, \vv, s, \DD,\BB)$, we have to use material relations 
\begin{equation}\label{eq.mat}
\DD' = \eepsilon \cdot \EE'
\qquad\mbox{and}\qquad
\BB' = \mmu \cdot \HH',
\end{equation}
which are valid in the frame co-moving with the center of mass of the matter, i.e. with velocity $\vv$ with respect to the laboratory frame. Symmetric positive definite tensors $\eepsilon$ and $\mmu$ represent the electric permittivity and magnetic permeability. The co-moving fields $\DD'$, $\BB'$, $\EE'$ and $\HH'$ are then transformed to the laboratory fields by transformation \eqref{eq.trafo.Gal}. This leads to energy
\begin{equation}\label{eq.E.EMHD}
E_{EMHD} = \int\dd\rr \left(\frac{1}{2}\left(\sum_\alpha \rho_\alpha\right) \vv^2 
+ \frac{1}{2}\DD\cdot\eepsilon^{-1}\cdot\DD  + \frac{1}{2}\BB\cdot\mmu^{-1} \cdot \BB
+\vv\cdot (\DD\times\BB) + \eps(\rho_\alpha, s)\right).
\end{equation}
This energy has the right behavior with respect to Galilean transformations, see e.g. \cite{leblond1973galilean}.

The evolution equations implied by bracket \eqref{eMIX03} and energy \eqref{eq.E.EMHD} are
\begin{subequations}\label{eq.EMHD.v}
\begin{align}
 (\partial_t \rho_\alpha)_\text{rev} &= -\partial_i (\rho_\alpha v^i)\\
\label{eMIX16}\rho (\partial_t v_i)_\text{rev}& = %
 -\rho v^j\partial_j v_i -\sum_\alpha \rho_\alpha\partial_i \frac{\partial \eps}{\partial \rho_\alpha} - s\partial_i \frac{\partial \epsilon}{\partial s} \nonumber\\
 &+ \sum_{\alpha=1}^n e_0\frac{z_\alpha \rho_\alpha}{m_\alpha}\underbrace{\left(E_i  + \varepsilon_{ijk} v^j B^k  \right)}_{= D_i^\dagger}~\\
(\partial_t s)_\text{rev} &= -\partial_i (s v^i)\\
\label{eq.D.EMHD.v}(\partial_t D^i)_\text{rev} &= \epsilon^{ijk}\partial_j H_k -\sum_\alpha e_0\frac{z_\alpha \rho_\alpha}{m_\alpha}v^i\\
\label{eq.B.EMHD.v}(\partial_t B^i)_\text{rev} &= -\epsilon^{ijk}\partial_j E_k,
\end{align}
where we took velocity $\vv$ as a state variable instead of the mass momentum $\uu=\rho\vv$, $\rho=\sum_\alpha\rho_\alpha$ being the total density. The conjugate quantities remain the same because derivative of energy keeping $\uu$ and $\rho$ constant is the same as when keeping $\vv$ and $\rho$ constant. Note that $v_i = g_{ij}v^j$ is the velocity multiplied by the Eucledian metric (equal to the unit matrix).
\end{subequations}
The terms on the last line represent the Lorentz force acting upon the free charge density. Equations \eqref{eq.EMHD.v} are invariant with respect to Galilean transformations \eqref{eq.trafo.Gal}. Therefore, they represent a Galilean invariant form of electrodynamics coupled with matter. Note that Galilean invariance would be impossible without the explicit presence of evolution equations for matter.

The relation between formulation~\eqref{eMIX03} and~\eqref{eMIX05} can also be understood in terms of surface balances of the electric induction flux and magnetic induction flux for static and moving surfaces as it is shown in~\cite[Eqns.\ 2.95, 2.96]{marsik1998biotermodynamika}

Note that the bracket formulated with the total momentum~\eqref{eMIX05} satisfies the Jacobi identity unconditionally. This is not the case for the bracket with the mass momentum~\eqref{eMIX03}, where the validity of the Jacobi identity cannot be established without $\dive\BB = 0$.

\subsubsection{Gau{\ss}'s law for electric charge and non-existence of magnetic monopoles}
Equations~\eqref{VED04} represent the Gau{\ss}'s laws.
Let us now consider the dynamics of coupled matter and electromagnetic field generated by brackets~\eqref{eMIX03} and~\eqref{eMIX05}. After applying divergence on Eq. \eqref{eq.D.EMHD.v} we obtain that 
\begin{align}
    \label{GLEM07}  \dive \DD = \rho_\textrm{f} \stackrel{def}{=} \sum_\alpha e_0\frac{z_\alpha \rho_\alpha}{m_\alpha},
\end{align}
where $\rho_f$ stands for the \textbf{free charge}. 
The equality holds if it holds at some initial time (which can be for instance vacuum, so it indeed holds). Applying divergence on Eq. \eqref{eq.B.EMHD.v}, we obtain again that $\dive\BB=0$. The Gau{\ss}'s law is implied by the evolution equations even in the case of electrodynamics coupled with mixtures of fluids.

\subsection{Polarization}\label{sec.pol}
The reversible evolution of a charged mixture in electromagnetic field is described by one of the Poisson brackets in the previous section and a choice of energy.
But such description does not, in general, capture the intrinsic dipole moments of the molecules, i.e. \textit{polarization}.
Indeed, a fluid of dipoles can be charge neutral while still electromagnetically interacting. 
An additional \textit{bound} charge is present due to internal dipole density of the matter on top of the modeled free charge.

Description of the bound charge depends profoundly on the chosen variables, the time/space scales and the internal structure of the assumed matter.
The classical treatment on the macroscopic level, see e.g.~\cite{purcell1985electricity}, resorts to the definition of polarization vector $\PP$.
The divergence part of $\PP$ is set equal to the density of bound charge.
The time derivative of $\PP$ represents current, therefore it is added to the left-hand side of Ampere's law \eqref{eq.D.EMHD.v}.

Russakoff in~\cite{russakoff1970derivation} acquired the polarization as a consequence of averaging of microscopic Maxwell's equations with point charges and subsequent expansion with respect to spatially correlated charges.
This approach leads, compared to the Purcell's, to a definition of the polarization related with magnetization proportional to the averaged relative velocities of the correlated charges.

\subsubsection{Intrinsic dynamics of polarization}
Density of polarization $\PP$ represents a vector field just as the displacement field $\DD$. Advection of the vector field by the fluid mechanics is then expressed by a semidirect product as in Sec. \ref{sec.SP}. Similarly as in that Section, we can add (besides $\PP$) the conjugate momentum variable (to be denoted by $\ppi$). The Poisson bracket expressing advection of the pair $(\ppi,-\PP)$ by fluid mechanics is shown in the following section.

The canonical Poisson bracket~\eqref{cpb} of state variables $\ppi$ and $\PP$ reads
\begin{align}
    \{F,G\}^{\PP\ppi} (\PP,\ppi) = \int\dd\rr\left( F_{P^i} G_{\pi_i}-G_{P^i} F_{\pi_i} \right).
    \label{HPM01}
\end{align}
Bracket~\eqref{HPM01} can represent a continuum of elementary dipoles with fixed centers of mass, but changing length and orientation.
Indeed, covector field $\ppi$ can be interpreted as proportional to the relative momentum of particles forming the dipole, c.f.\ variable~$\textbf{t}$ in~\eqref{EDI04} and~\eqref{EDI08}.
The divergence part of $\PP$ represents the bound charge density, 
\begin{align}
	\rho_b \stackrel{\mbox{def}}{=} -\dive \PP.
  \label{HPM08}
\end{align}

\subsubsection{Advected cotangent bundle $(\ppi,-\PP)$}
When the dipoles are not fixed in space, but advected by a fluid, the
interaction is captured by coupling of bracket~\eqref{HPM01} and fluid dynamics bracket~\eqref{CIT01} by semidirect product,
\begin{align}
\label{HPM02}%
\left\{F,G\right\}^{\textrm{FMP}\ppi}(\rho_\alpha,& \umu\uu, s, \PP, \ppi)=\{F,G\}^{\textrm{(CIT)}}(\rho_\alpha, \umu\uu,s) + \{F,G\}^{\PP\ppi}(\PP,\ppi)\nonumber\\
&\phantom{--}+\int\dd\rr P^i\left(\partial_j F_{P^i} G_{\umu u_j} - \partial_j G_{P^i}F_{\umu u_j}\right)\nonumber\\
&\phantom{--}+\int\dd\rr \partial_j P^j(F_{\umu u_i} G_{P^i} -G_{\umu u_i} F_{P^i})\nonumber\\
&\phantom{--}+\int\dd\rr P^j(  F_{\umu u_i} \partial_j G_{P^i} - G_{\umu u_i} \partial_j F_{P^i})\\
\{F,G\}^{\textrm{(SP)}_\AA}(\ppi,\umu\uu)&
\begin{dcases}
&-\int\dd\rr \partial_j\pi_i\left(F_{\pi_i} G_{\umu u_j} -G_{\pi_i} F_{\umu u_j} \right)\nonumber\\
&-\int\dd\rr \pi_i\left(F_{\pi_j}\partial_j  G_{\umu u_i} - G_{\pi_j}\partial_j  F_{\umu u_i} \right)\nonumber,
\end{dcases}
\end{align}
see bracket \eqref{Can-LP-coupling} later comments for more details. This Poisson bracket expresses kinematics of state variables $(\rho_\alpha,\umu\uu,s,\PP,\ppi)$.
Note that the total momentum of the coupled system is denoted as $\umu\uu$. This Poisson bracket expresses kinematics of fluid mechanics advecting the polarization density with its conjugate momentum (relative momentum of the intrinsic dipole charges).

Note that bracket~\eqref{HPM02} may be briefly expressed as
\begin{align}
    \left\{F,G\right\}^{\textrm{FMP}\ppi}(\rho_\alpha,& \umu\uu, s, \PP, \mmu)=\{F,G\}^{\textrm{(CIT)}}(\rho_\alpha, \umu\uu,s) + \{F,G\}^{\PP\mmu}(\PP,\ppi)\nonumber\\
    &+\{F,G\}^{\textrm{(SP)}}(\PP,\umu\uu) + \{F,G\}^{\textrm{(SP)}_\AA}(\ppi,\umu\uu)~,
\label{HPM06}
\end{align}
using the definition of $\{F,G\}^{\textrm{(SP)}}$ from~\eqref{eMIX05} and $\{F,G\}^{\textrm{(SP)}_\AA}$ from~\eqref{HPM02}.

The evolution equations given by~\eqref{HPM02} are as follows,
\begin{subequations}
\label{HPM05} 
\begin{align}
  (\partial_t \rho_\alpha)_\text{rev}%
  &= - \partial_j(\rho_\alpha \umu u^{\dagger j})\\
  (\partial_t s)_\text{rev}%
  &= - \partial_j(s \umu u^{\dagger j})\\
  (\partial_t \umu u_i)_\text{rev}%
  & = -\sum_{\alpha=1}^n\rho_\alpha\partial_i \rho_\alpha^\dagger -s\partial_i s^\dagger -\umu u_j\partial_i \umu u^{\dagger j} 
       -P^j\partial_i P_j^\dagger +\pi^{\dagger j}\partial_i\pi_j \nonumber\\
  &+ \partial_j\left(
      P_i^\dagger P^j -  \pi_i\pi^{\dagger j} - \umu u_i \umu u^{\dagger j}
    \right)~,\\
  (\partial_t P^i)_\text{rev}%
  &= \pi^{\dagger i} - \partial_j\left( P^i \umu u^{\dagger j} - P^j \umu u^{\dagger i} \right) - \umu u^{\dagger i}\partial_j P^j \label{HPM05d}\\
  (\partial_t \pi_i)_\text{rev}%
  &= -P_i^\dagger - \umu u^{\dagger j}\partial_j \pi_i  - \pi_j \partial_i \umu u^{\dagger i}.
\end{align} 
Later they will be equipped with dissipation of $\ppi$ triggering subsequent relaxation of polarization.
\end{subequations}

Note that the semi-direct product theory does not fully determine the relation of the total momentum $\umu\uu$ and the mass momentum. We only know that the velocities coincide, which means that
\begin{align}    
\rho\left(\umu\uu\right)^\dagger = \rho\mathbf{v} = \uu~.
\label{HPM07}
\end{align}
This relation can be ascertained by further assumptions on the physical properties of the interaction.

Let us assume $\umu\uu = \uu$, i.e., there appears no extra momentum on top of the mass momentum.
The assumption corresponds with the nature of the transformation~\eqref{EDI02}, where motion of an elementary dipole is described by two de-coupled canonical brackets, one for the center of mass and one for the relative quantities, and the center-of-mass momentum of the dipole affects neither $\PP$ nor $\ppi$ directly.

\subsubsection{Coupling to the electromagnetic field -- total charge density}
Having coupled fluid mechanics with polarization density and its conjugate momentum, let us finally make the coupling to the electromagnetic field, i.e. having state variables $(\rho_\alpha,\umu\uu,s, \DD,\BB,\PP,\ppi)$. 
Both pairs $(\ppi,\PP)$ and $(\DD,\BB)$ were coupled to fluid mechanics by semidirect product. Advection of both pairs by fluid mechanics can be thus expressed (using ~\eqref{VED02},~\eqref{eMIX05},~\eqref{HPM02})  by Poisson bracket
\begin{align}
\label{CE01}
\{F,G\}^{\textrm{EFP}\ppi}(\rho_\alpha,{}&\umu\mm,s, \DD, \BB,\PP,\ppi)=\{F,G\}^{\textrm{(CIT)}}(\rho_\alpha,\umu\mm,s)\\
{}&+ \{F,G\}^\text{(EM)}(\DD, \BB) + \{F,G\}^\text{(SP)}(\umu\mm, \DD)+ \{F,G\}^{\text{(SP)}}(\umu\mm, \BB)\nonumber\\
{}&+ \{F,G\}^{\PP\ppi}(\PP, \ppi)+ \{F,G\}^\text{(SP)}(\umu\mm, \PP) + \{F,G\}^{\text{(SP)}_\AA}(\umu\mm, \mmu),\nonumber
\end{align}
which, however, does not contain any direct coupling between $\PP$ and $\DD$.
The coupling appears after the transformation to the field of \textit{electric induction} $\bcDD$,
\begin{align}
    \label{TC01}
    \bcDD = \DD - \PP,
\end{align}
for which it holds that
\begin{align}
    \dive \bcDD = \dive\DD - \dive\PP =\rho_\textrm{f} +\rho_\textrm{b}~, %
    \label{TC09}
\end{align}
where the second equality is due to~\eqref{HPM08} and~\eqref{GLEM07}. 
The divergence of $\bcDD$ represents the total charge density and is given as a sum of the bound charge density and the free charge density.

For a functional $\widehat{F}(\DD, \PP, \dots) = F(\bcDD,\PP, \dots)$ then holds that
\begin{align}
\label{TC10}
	\widehat{F}_\PP = F_\PP - F_{\bcDD}\quad\mbox{and}\quad \widehat{F}_\DD = F_{\bcDD}~.
\end{align}
Poisson bracket~\eqref{CE01} can be reformulated using the electric induction~\eqref{TC01} and the transformation~\eqref{TC10} as
\begin{align}
\label{TC11}
    \{F,G\}^{\bcDD\ppi}(\rho_\alpha,{}&\umu\mm,s, \cDD, \BB,\PP,\mmu)=\nonumber\\
    {}&\{F,G\}^{\textrm{EFP}\ppi}(\rho_\alpha,\umu\mm,s, \bcDD, \BB,\PP,\ppi)- \{F,G\}^{\PP\ppi}(\bcDD, \ppi)~.
\end{align}
Surprisingly, just the evolution equations of $\bcDD$ and of polarization momentum $\ppi$ are affected by the transformation.
The transformed evolution equations of the electric induction and polarization momentum $\ppi$ then become
\begin{subequations}
\begin{align}
(\partial_t\cDD^i)_\text{rev}=%
      {}&\varepsilon_{ijk}\partial_j B_k^\dagger - \pi^{\dagger i} - \partial_j\left(\cDD^i v^j - \cDD^j v^i \right) -  v^i\partial_j \cDD^j~,\label{TC02}\\
(\partial_t\pi_i)_\text{rev}=%
      {}& -P_i^\dagger + \cDD_i^\dagger - \partial_j\pi_i v^j- \pi_j \partial_i v^j ~.\label{TC05}
\end{align}
\end{subequations}
Note that the Gau{\ss}'s law \eqref{TC09} is again compatible with the evolution of the $\dive\bcDD$, given by~\eqref{TC02}.

The evolution of the bound charge density, given by bracket~\eqref{TC11} and formally identical to~\eqref{HPM05d}, reads
\begin{align}
    (\partial_t \partial_i P^i)_\text{rev} = \partial_i \pi^{\dagger i} - \partial_i \left(v^i \partial_j P^j  \right)~.
    \label{TC20}
\end{align}
The divergent part of~\eqref{TC20} evolves as a convected density which also be changed by $\ppi^\dagger$. The term $\ppi^\dagger$ can be interpreted as a reversible polarization current. This reversible current appears in the evolution of $\bcDD$, see Eq. ~\eqref{TC02}.

The material relation for electric intensity now changes to 
\begin{align}
    \label{TC07}
   \eepsilon_0\cdot \EE' = \DD' - \PP' ~, 
\end{align}
which is the usual relation in the frame co-moving with matter. Transformation \eqref{eq.trafo.Gal} brings these relation to the laboratory frame.

\paragraph{Total momentum.}
In order to obtain the Lorentz force, the total momentum $\umu\mm$ has to be related to the mass momentum $\uu$. We choose
\begin{align}
    \umu\mm = \uu + \bcDD\times\BB~,
    \label{TC13}
\end{align}
since it is compatible with the choices made for the polarization-fluid coupling~\eqref{HPM07} and the EMHD coupling~\eqref{eMIX04}. The bracket~\eqref{TC11} may be transformed using~\eqref{TC13} and transformation rules analogous~\eqref{eMIX17} so that the description with the mass momentum $\uu$ is obtained. The calculation is analogous to the transformation in the case of EMHD brackets~\eqref{eMIX05} to~\eqref{eMIX03}. Apart from the presence of $(\PP, \ppi)$ and its coupling to the fluid mechanics, the only actual difference is the $\{\cdot,\cdot\}^{\PP\ppi} (\bcDD, \ppi)$ appearing in~\eqref{TC11}. 
Relation~\eqref{TC13} implies the following form of the Lorenz force density (right hand side of the evolution equation for momentum $\uu$)
\begin{align}
    \label{TC15}
    \partial_j \cDD^j \left(  \cDD_i^\wdagger + \varepsilon_{ijk} v^j B^k \right)
    +\varepsilon_{ijk} \pi^{\wdagger j} B^k~,
\end{align}
where $\xx^\wdagger$ denotes derivative of energy dependent on $\uu$ with respect to $\xx$, analogously as in Section~\ref{sec:mass_momentum} transformation~\eqref{eMIX17}.

Let us analyze~\eqref{TC15}. The first term contains the electric field $\EE=\bcDD^\wdagger$ and $\vv\times\BB$ acting upon the total charge density $\dive\bcDD$. The main difference to the Lorentz force appearing in~\eqref{eMIX16}, is the total charge density appearing in front of the bracket.
The second term in~\eqref{TC15} is a product of the magnetic field $\BB$ and the reversible rate of change of polarization $\ppi^\wdagger$.
Hence, it can be interpreted as a force exerted by the magnetic field due to the current of the polarized charge.
Due to the form of~\eqref{TC13} it holds $\ppi^{{\dagger}}= \ppi^\wdagger$ and $\PP^{{\dagger}}= \PP^\wdagger$.

Energy \eqref{eq.E.EMHD} turns to
\begin{multline}\label{eq.E.EMHDPmu}
E_{EMHDP\ppi} = 
\int\dd\rr \left(\frac{1}{2}\left(\sum_\alpha \rho_\alpha\right) \vv^2 
+ \frac{1}{2}\bcDD\cdot\eepsilon^{-1}\cdot\bcDD  + \frac{1}{2}\BB\cdot\mmu^{-1} \cdot \BB\right.\\
\left.\vphantom{\int}+\vv\cdot (\bcDD\times\BB) + \frac{1}{2}\PP\cdot\frac{\eepsilon_0^{-1}}{\chi}\cdot\PP + \eps(\rho_\alpha, s, \PP, \ppi) \right).
\end{multline}
Note that for pure substances the tensor $\eepsilon$ in this formula for energy becomes the vacuum permittivity (multiplied by the inverse metric) because then the usual formula for dielectrics is recovered, see Sec. \ref{sec.dielectrics}, after the polarization has relaxed to equilibrium. However, for a mixture one should consider a polarization for each constituent, and some of the polarizations can be relaxed while some not. We keep the tensor $\eepsilon$ in the formula for energy in order to cover also the mixtures.

In summary, reversible evolution equations for a mixture coupled with electromagnetic fields, polarization and its conjugate momentum were constructed in a geometric way (semidirect products).

\subsubsection{Polarization waves and stress}
Let us demonstrate the dynamics of the $\PP$, $\ppi$ couple on the canonical polarization bracket~\eqref{HPM01}, bringing new equations for propagation of polarization. For the sake of simplicity, we assume state variables $(\PP,\ppi)$ in this Section.
Let us assume that the energy contains a contribution corresponding to a dipole-dipole interaction that is proportional to the square of the polarization gradient $\nabla\PP$.
Hence, we have
\begin{align}
    E^\textrm{Pw} = \int\dd\rr\frac{1}{2} \left( \beta_{ij}\partial_j P^i \partial_j P^i + \alpha^ij  \pi_i \pi_j\right)~,
    \label{PW01}
\end{align}
where $\alpha^{ij}$ and $\beta_{ij}$ are assumed to be symmetric positive definite constant tensors.
The Hamiltonian evolution of $\PP$ and $\ppi$ due to bracket~\eqref{HPM01} with energy~\eqref{PW01} reads
\begin{subequations}
    \label{PW02}
\begin{align}
    (\partial_t P^i)_\text{rev}    &= \alpha^{ij}\pi_j\\
    (\partial_t \pi_i)_\text{rev}  &= \beta_{ik}\partial_j\partial_j P^k.
\end{align}
\end{subequations}
Assuming isotropy, $\alpha^{ij} = \alpha g^{ij}$ and $\beta_{ij} = \beta g_{ij}$, where $\alpha$ and $\beta$ are positive numbers and $g_{ij}$ the Eucledian metric, Eqs. \eqref{PW02} become
\begin{align}
    (\partial^2_{tt} P^i)_\text{rev}    &= \alpha(\partial_t\pi_i)_\text{rev} = \alpha\beta\partial_j\partial_j P^i~,
\end{align}
which is a polarization wave equation. Speed of the polarization sound is $(\alpha\beta)^{-1/2}$. 

Let us now consider the level of description with state variables $(\rho_\alpha, s, \umu\mm, \cDD, \BB, \PP, \ppi)$, which is equipped with bracket \eqref{CE01} and where~\eqref{PW01} contributes to the energy.
The evolution equation for total momentum $\umu\mm$ is given by the bracket~\eqref{TC11} also determines the evolution of the mass momentum using transformation~\eqref{TC13}. Thus we have 
\begin{align}
(\partial_t u_i)_\text{rev} = %
	&-\sum_{\alpha=1}^n \rho_\alpha\partial_i \rho_\alpha^\wdagger -s\partial_i s^\wdagger -u_j\partial_i u_j^\wdagger 
	- \partial_j\left(u_i u_j^\wdagger\right) \nonumber\\
          & \underbrace{+\beta P_j\partial_i\partial_k\left(\partial_k P^j \right) +\alpha\pi_j\partial_i\pi_j 
          + \partial_j\left(- \beta \partial_k\partial_k P^i P^j - \alpha \pi_i \pi_j \right)}_{\text{Dipole-dipole interaction}}\nonumber\\
    &\underbrace{+\partial_j \cDD^j \left(  \cDD_i^\wdagger + \varepsilon_{ijk} v_j B^k \right) 
    +\varepsilon_{ijk} \alpha\pi_j B^k}_{\text{Lorentz force}}~,
\label{PW03}
\end{align}
without specifying the energy dependence of the other variables. Note that we no longer pay attention the upper and lower indexes in this section for simplicity.
The last line is the Lorentz force acting upon the total charge, c.f.~\eqref{TC15}.
The terms with $\PP$ and $\ppi$ on the second line of the right hand side of~\eqref{PW03} can be written in a divergence form as
\begin{align}
    \label{PW04}
  \partial_j \left(
      \beta P_l \partial_i( \partial_j P_l) -\delta_{ij} \frac{\beta}{2}\partial_k P_l \partial_k P_l 
    - \beta\partial_k(\partial_k P^i)P^j -\alpha\pi_i\pi_j +\frac{\alpha}{2}\delta_{ij}\pi_k\pi_k
  \right)~, 
\end{align}
which is the stress contribution due to the dipole-dipole interaction. Momentum is conserved.

The third term in~\eqref{PW04} is not symmetric w.r.t.\ change of $i$ and $j$, but conservation of the angular momentum is not violated.
Indeed, the cross product of the discussed term and the position vector $\rr$ can be written in a divergence form. We have, omitting $\beta$,
\begin{align}
 \varepsilon_{sri} r_r\partial_j& \left(P^j \partial_k(\partial_k P^i) \right) 
= \varepsilon_{sri}\partial_j \left( r_r P^j \partial_k( \partial_k P^i) \right) 
- \varepsilon_{sri}P^j \partial_k( \partial_k P^i) \delta_{rj}\nonumber\\
&=\varepsilon_{sri}\partial_j \left( r_r P^j \partial_k( \partial_k P^i) \right) 
- \varepsilon_{sji} \partial_k(P^j \partial_k P^i) 
+ \varepsilon_{sji} \partial_k P^j \partial_k P^i ~,
    \label{PW05}
\end{align}
where the last term is simultaneously symmetric and antisymmetric in $i$ and $j$, hence it is zero. Similar observation was made in \cite{extra-mass-flux}. The angular momentum is conserved for the proposed energy of the dipole-dipole interaction~\eqref{PW01}.

\subsection{Magnetization}\label{sec.mag}

The magnetization of matter, see e.g.~\cite[Sec.\,11]{purcell1985electricity}, is due to the orientation of spins. Perhaps due to the resemblance with dynamics of rigid body rotation, the pioneering model of magnetization by Landau \& Lifshitz \cite{LL1935} was based on that dynamics.
In the following text we first recall the Hamiltonian formulation of rigid body dynamics and then we let the rigid body dynamics be advected by the fluid (using again the semidirect product theory).

\subsubsection{Intrinsic dynamics of magnetization}
The configuration manifold of rigid body rotations is the Lie group $SO(3)$. The standard machinery of differential geometry, e.g. \cite{arnold,Fecko} or ~\cite[Eq.\,3.69]{pavelka2018multiscale}, concludes that the Lie algebra dual, where angular momentum $\MM$ seen from the body reference frame plays the role of state variable, is equipped with a Poisson bracket. Local version of the bracket (see also \cite{dv}) is
\begin{align}
    \{F,G\}^{\text{SO(3)}}&{}(\MM)= -\gamma \int\dd\rr M_i\varepsilon_{ijk} F_{M_j} G_{M_k}~,
    \label{LL10}
\end{align}
where $\gamma$ is the gyromagnetic ratio and $\MM(\rr)$ is the local angular momentum covector field.
Bracket~\eqref{LL10} is also called the spin bracket, see e.g.~\cite{lakshmanan2011fascinating}. The evolution equation implied by this bracket is (by the same procedure as in Sec. \ref{sec.FM})
\begin{equation}
    (\dot{\MM})_\text{rev} = \gamma\MM \times E_{\MM},
\end{equation}
$E$ being energy of the rotation. Derivative of energy with respect to $\MM$ is the angular velocity $\oomega$.

\subsubsection{Advection by fluid mechanics}
Similarly as in the case of electromagnetic field or polarization, we will now construct the Poisson bracket expressing advection of magnetization and its dynamics by fluid mechanics (using again semidirect product). Now, however, the advected structure is not a cotangent bundle, but a Lie algebra dual, i.e. having state variables $(\rho, \uMM\mm, s, \MM)$.
The general formula for semidirect product of two Lie algebra duals, see ~\cite[Eq. 40]{esen2017hamiltonian}, then gives Poisson bracket
\begin{align}
    \{F,G\}^{\text{SO(3)}\times\text{(mEHMD)}}&{}(\rho_\alpha, \uMM\mm, s, \DD, \BB, \MM)= \{F,G\}^\textrm{mEHMD} + \{F,G\}^\textrm{SO(3)}\nonumber\\
    &{}+ \left\langle \MM, F_{\uMM\mm}\rhd G_\MM\right\rangle - \left\langle \MM, G_{\uMM\mm}\rhd F_\MM\right\rangle~,
    \label{LL01}
\end{align}
where $\uMM\mm$ is the new total momentum  of the coupled system and $\MM$ denotes the magnetization. Note that the advected electrodynamics are kept in the bracket for completeness.
The right action of velocity field $F_\mm$ on the vector field $G_\mm$ is defined as negative of Lie derivative as usually,
\begin{align} 
    \left(F_{\uMM\mm} \rhd G_\MM\right)_i = -\left( F_{\uMM m_j}\partial_j G_{M_i} - G_{M_j}\partial_j F_{\uMM m_i}\right)~. 
    \label{LL02}
\end{align}
Using~\eqref{LL02}, bracket ~\eqref{LL01} becomes
\begin{align}
    \{F,G\}^{\text{SO(3)}\times\text{(mEHMD)}}&{}(\rho_\alpha, \uMM\mm, s, \DD, \BB, \MM) = \{F,G\}^\textrm{mEHMD} + \{F,G\}^\textrm{SO(3)}\nonumber\\
    {\{F,G\}^{\text{SP}_\MM}(\uMM\mm,\MM):=}&
\begin{dcases}
   &{}+\int\dd\rr M_i\left(\partial_j F_{\uMM m_i} G_{M_j} - \partial_j G_{\uMM m_i} F_{M_j}  \right)\nonumber\\
   &{}+\int\dd\rr M_i\left(\partial_j F_{M_i} G_{\uMM m_j} - \partial_j G_{M_i} F_{\uMM m_j}  \right),
\end{dcases}\\
    \label{LL03}
\end{align}
which is the explicit form of Poisson bracket expressing kinematics of magnetization advected by fluid mechanics.

Let us choose  
\begin{align}
    \uMM\mm = \uu + \DD\times\BB~,
    \label{LL05}
\end{align}
which, as in the polarization case, means that there is no excess momentum and $\uMM\mm = \mm$.

The evolution equation equations implied by~\eqref{LL03} are
\begin{subequations}
\label{LL04}
\begin{align}
  (\partial_t  m_i)_\textrm{rev} %
  & = -\sum_{\alpha=1}^n\rho_\alpha\partial_i \rho_\alpha^\dagger -s\partial_i s^\dagger - m_j\partial_i  m^{\dagger j}
       -D^j\partial_i D_j^\dagger -B^j\partial_i B_j^\dagger - M_j\partial_i M^{\dagger j} \nonumber\\
  &+ \partial_j\left(
      D_i^\dagger D^j + B_i^\dagger B^j - M_i M^{\dagger j} -  m_i  m^{\dagger j}
    \right)~,\label{LL04a}\\
  (\partial_t M_i)_\textrm{rev} %
  &= \gamma \varepsilon_{ijk} M_j M^{\dagger k} - M_j\partial_i  m^{\dagger j}  - \partial_j\left( M_i  m^{\dagger j}\right)~.\label{LL04b}
\end{align}
\end{subequations}
These evolution equations show how magnetization is advected by fluid mechanics, and how such advection affects the fluid motion itself. Moreover, magnetization keeps its intrinsic rigid-body-like dynamics. Finally, note that there is no explicit coupling to the electromagnetic field just as in the original \cite{LL1935} paper. The coupling is achieved implicitly later by letting energy depend on both $\MM$ and $\BB$.

\subsubsection{Spin-spin interaction stress}
Let us now suggest novel dynamics of magnetization advected by fluid mechanics.
Let now $(\rho_\alpha, \mm, s, \DD, \BB, \MM)$ be the considered level of description governed by~\eqref{LL01}. 
The spin-spin interaction contribution to the energy can, on the continuum level, be modeled by the gradient of magnetization $\nabla\MM$ as in~\cite{LL1935},
\begin{align}
    E^\textrm{Ms} = \int\dd\rr \frac{\alpha^{ik}}{2} \partial_j M_i \partial_j M_k~,
    \label{MSI01}
\end{align}
where $\alpha^{ik}$ is a constant positive definite tensor.
The mass momentum $\uu$, given by the evolution of the total momentum~\eqref{LL04a}, relation~\eqref{LL05} and relations analogous to~\eqref{eMIX15}, evolves as 
\begin{align}
(\partial_t u_i)_\text{rev} =%
  &  -\sum_{\alpha=1}^n\rho_\alpha\partial_i \rho_\alpha^\dagger -s\partial_i s^\dagger -u_j\partial_i u^{\dagger j} - \partial_j\left(u_i u^{\dagger j} \right)
       \nonumber\\
  &+ \gamma \alpha M_j\partial_i \partial_k \partial_k M_j+ \partial_j\left(
       \gamma \alpha M_i \partial_k \partial_k M_j 
   \right)\nonumber\\
   & +\partial_j D^j \left(  D_i^\wdagger + \varepsilon_{ijk} u^{\dagger j} B^k \right)
    ~,\label{MSI03}
\end{align}
where only the derivatives of energy w.r.t.\ $\MM$ are expressed explicitly. Note that we have put $\alpha^{ij}=\alpha g^{ij}$ and that we stopped distinguishing between upper and lower indexes for simplicity. The terms on the last line represent the Lorentz force acting on the free charge. The terms on the second line in~\eqref{MSI03} are the stress due to the spin-spin interaction.
The terms can be, similarly to the dipole-dipole interaction~\eqref{PW04}, written in the divergence form as
\begin{align}
    \gamma\alpha\partial_j \left( M_l\partial_i\partial_j M_l - \frac{\delta_{ij}}{2}\partial_k M_l \partial_k M_l +  M_i \partial_k \partial_k M_j  \right)~.
    \label{MSI04}
\end{align}
The non-symmetric term in~\eqref{MSI04} is equivalent to the one already discussed in the polarization case~\eqref{PW05}, therefore, it also does not violate the conservation of the angular momentum.

\subsection{General level $(\rho_\alpha, \widehat\mm, s,\cDD,\BB,\PP,\ppi,\MM)$}
The hierarchy of the brackets built in the preceding paragraphs will be completed on the level of description containing all the discussed variables.
The General bracket reads
\begin{align}
\label{GG01}
\{F,G\}^{\textrm{GEMHD}}(\rho_\alpha,{}&\widehat\mm,s, \bcDD, \BB,\PP,\ppi, \MM)=\{F,G\}^{\textrm{CIT}}(\rho_\alpha, \widehat\mm, s)\nonumber\\
{}&+ \{F,G\}^\text{(EM)}(\bcDD, \BB) + \{F,G\}^\text{(SP)}(\widehat\mm, \bcDD)+ \{F,G\}^{\text{(SP)}}(\widehat\mm, \BB)\nonumber\\
{}&+ \{F,G\}^{\PP\ppi}(\PP, \ppi)\phantom{+}+ \{F,G\}^\text{(SP)}(\widehat\mm, \PP) + \{F,G\}^{\text{(SP)}_\AA}(\widehat\mm, \ppi)\nonumber\\
{}&+ \{F,G\}^{\text{SO(3)}}(\MM)\phantom{+} + \{F,G\}^{\text{(SP)}_\MM}(\widehat\mm,\MM)~.
\end{align}
The first line is due to the dynamics of the CIT-mixture.
The second line of~\eqref{GG01} accounts for the electromagnetism and its coupling to continuum. 
The third line contains the polarization bracket and its coupling to continuum.
The fourth line of~\eqref{GG01} is due to the magnetization dynamics and its coupling to continuum.
Brackets $\{\cdot,\cdot\}^\textrm{(SP)}$, $\{\cdot,\cdot\}^{\textrm{(SP)}_\AA}$ and $\{\cdot,\cdot\}^{\textrm{(SP)}_\MM}$ were found due to the semidirect product theory.

The evolution equations implied by the General bracket are
\begin{subequations}
\label{CG02}
\begin{align}
  (\partial_t \rho_\alpha)_\text{rev}%
  &= - \partial_j(\rho_\alpha \widehat m_j^\dagger)\\
  (\partial_t s)_\text{rev}%
  &= - \partial_j(s \widehat m_j^\dagger)\\
  (\partial_t \widehat m_i)_\text{rev}%
  & = -\sum_{\alpha=1}^n\rho_\alpha\partial_i \rho_\alpha^\dagger -s\partial_i s^\dagger -\widehat m_j\partial_i {\widehat m_j}^\dagger \nonumber\\
       & -\cDD^j\partial_i \cDD_j^\dagger -B^j\partial_i B_j^\dagger -P^j\partial_i P_j^\dagger -\pi_j\partial_i\pi^\dagger_j- M_j\partial_i M_j^\dagger \nonumber\\
  &+ \partial_j\left(
	\cDD_i^\dagger \cDD^j +B_i^\dagger B^j + P_i^\dagger P^j -  \pi_i\pi^{\dagger j}+\pi_k \pi^\dagger_k\delta_{ij} - M_i M_j^\dagger- \widehat m_i {\widehat m_j}^\dagger
    \right)~,\\
  (\partial_t \cDD^i)_\text{rev}%
  {}&=\varepsilon_{ijk}\partial_j B_k^\dagger - \pi^{\dagger i} - \partial_j\left(\cDD^i \widehat m_j^\dagger - \cDD^j \widehat m_i^\dagger \right) -  \widehat m_i^\dagger\partial_j \cDD^j~,\\
  (\partial_t B^i)_\text{rev}%
  &= -\varepsilon_{ijk}\partial_j \cDD_k^\dagger - \partial_j\left( B^i {\widehat m_j}^\dagger - B^j {\widehat m_i}^\dagger \right) - {\widehat m_i}^\dagger\partial_j B^j \\
  (\partial_t P^i)_\text{rev}%
  &= \pi^{\dagger i} - \partial_j\left( P^i {\widehat m_j}^\dagger - P^j {\widehat m_i}^\dagger \right) - {\widehat m_i}^\dagger\partial_j P^j\label{GG02P} \\
  (\partial_t \pi_i)_\text{rev}%
  &= -P_i^\dagger + \cDD_i^\dagger - \partial_j\pi_i \widehat m_j^\dagger- \pi_j \partial_i \widehat m_j^\dagger\\
  (\partial_t M_i)_\textrm{rev} %
  &= \gamma \varepsilon_{ijk} M_j M_k^\dagger - M_j\partial_i \widehat m_j^\dagger  - \partial_j\left( M_i \widehat m_j^\dagger\right)~.
\end{align}
This is the most detailed set of reversible evolution equations expressing evolution of a mixture coupled with electromagnetic field, polarization and its conjugate momentum and magnetization.
\end{subequations}
When energy is among the state variables (thus transforming first to $\uu=\rho\vv$ to $\vv$), the evolution equations read
\begin{subequations}
\label{eq.gen.v}
\begin{align}
  (\partial_t \rho_\alpha)_\text{rev} &= - \partial_j(\rho_\alpha v^j)\\
  (\partial_t s)_\text{rev} &= - \partial_j(s v^j)\\
\label{eq.u.EMHDPmuM.v}
\rho (\partial_t v_i)_\text{rev}& = %
 -\rho v^j\partial_j v_i 
 + \partial_j \cDD^j\left(E_i  + \varepsilon_{ijk} v^j B^k  \right) + \varepsilon_{ijk}\pi^{\wdagger j}B^k~\nonumber\\
  &-\sum_\alpha \rho_\alpha\partial_i \frac{\partial \eps}{\partial \rho_\alpha} 
- s\partial_i \frac{\partial \epsilon}{\partial s} 
- P^j\partial_i \frac{\partial \epsilon}{\partial P^j}  
- \pi_j\partial_i \frac{\partial \epsilon}{\partial \pi_j}  
- M_j\partial_i \frac{\partial \epsilon}{\partial M_j}  \nonumber\\
  &+ \partial_j\left( 
    P_i^\wdagger P^j -  \pi_i\pi^{\wdagger j}+\pi_k \pi^{\wdagger k}\delta_{ij} - M_i M^{\wdagger j} 
  \right)~,\\
\label{eq.D.EMHDPmuM.v}
(\partial_t \cDD^i)_\text{rev} &= \epsilon^{ijk}\partial_j H_k -\sum_\alpha e_0\frac{z_\alpha \rho_\alpha}{m_\alpha}v^i- \pi^{\wdagger i}~,\\
 \label{GLEM01}
(\partial_t B^i)_\text{rev} &= -\epsilon^{ijk}\partial_j E_k~,\\
\label{eq.P.gen.v}
(\partial_t P^i)_\text{rev}  &= \pi^{\wdagger i} - \partial_j\left( P^i v^j - P^j v^i \right) - v^i \partial_j P^j~,\\
(\partial_t \pi_i)_\text{rev}  &= -P_i^\wdagger + (E_i+\eps_{ijk}v^j B^k) - \partial_j\pi_i v^j- \pi_j \partial_i v^j~,\\
\label{eq.M.gen.v}
  (\partial_t M_i)_\text{rev}  &= \gamma \varepsilon_{ijk} M^j M_k^\wdagger - M_j\partial_i v^j  - \partial_j\left( M_i v^j\right)~.
\end{align}
	Note that the form of energy is simpler when velocity $\vv$ is among the state variables instead of the momentum, as in Eq. \eqref{eq.E.EMHDPmu}. It should also be noted that these equations are not strictly Hamiltonian, since we do not have derivatives of energy w.r.t. velocity on the right hand side, but only velocity itself. The equations are equivalent to the Hamiltonian formulation \eqref{CG02}, but having velocity as a state variable is more suitable when formulating the energy.
\end{subequations}

The bracket~\eqref{GG01} can be projected to the levels of description upon which it was built.
One can simply evaluate the bracket~\eqref{GG01} on a set of functionals independent of a certain variables, see \cite{hierarchy}.
In the rest of this paper we enrich the reversible equations by irreversible terms in order to reduce this rather detailed description to the common continuum models coupling matter and electromagnetic field.

Consistently with the previously devised relations of the total momenta and mass momentum $\uu$, we choose
\begin{align}
    \widehat\mm = \umu\mm = \uu + \bcDD\times\BB~.
    \label{GG03}
\end{align}
The Lorentz force is equal to the one given by~\eqref{TC15}.

%
%
%
\section{Continuum thermodynamics and reductions}\label{sec:dissipation}
After having constructed a hierarchy of Poisson brackets for fluid mechanics of mixtures advecting electrodynamics, polarization and its conjugate momentum and magnetization, let us now enrich that detailed reversible dynamics by dissipative irreversible terms. This allows to see relaxation of fast mesoscopic variables and the effects on dynamics of less detailed variables. For instance we let the conjugate polarization momentum $\ppi$ relax to recover the standard Single Relaxation Time (SRT) model, which is widely used for comparison with experiments. We also let the magnetization $\MM$ dissipate to recover the full Landau \& Lifshitz model not only evolving in the laboratory frame, but being advected by the fluid. Finally, we approach the level of mechanical equilibrium, where neither momentum nor velocity is among the state variables. In particular, a generalized Nernst-Planck-Poisson equations are formulated on that level of description. In summary, a comprehensive multiscale thermodynamic construction of fluid mixtures equipped with electrodynamics, polarization and magnetization is provided.

\subsection{Gradient dynamics}
Before adding dissipative terms to the actual evolution equations, let us recall the general framework of gradient dynamics, where irreversible evolution is generated by derivatives of a dissipation potential \cite{Landau-Ginzburg,gyarmati}. Sound statistical arguments for gradient dynamics based on the large deviations principle was found in \cite{mielke2014,mielke-peletier,Alberto-FDT}.
The following paragraph closely follows~\cite[Sec. 4.5, 4.6]{pavelka2018multiscale}.

\subsubsection{Dissipation potential}
Consider a set of state variables $\xx$ and let energy, entropy and mass of the system be denoted by $E(\xx)$, $S(\xx)$ and $M(\xx)$, respectively.
A dissipation potential $\widetilde\Xi(\xx,\xx^*): \xx^* \rightarrow \mathbb{R}$ is a family of functionals of conjugate variables $\xx^*$ parametrized by $\xx$.
We require every parametrization $\Xi[\xx^*] = \widetilde\Xi(\xx)[\xx^*]$ to satisfy:
\begin{enumerate}[itemsep=3pt]
    \item Positiveness: $\Xi[\xx^*] \geq 0$ and $\Xi[\xx^*=0] = 0$.
    \item Monotonicity of derivative: $\left\langle \Xi_{\xx^*},\xx^*\right\rangle \geq 0\quad\forall\xx^*$.
    \item Convexity near $\xx^* = 0$.
    \item Degeneracy with respect to mass $\left\langle \Xi_{\xx^*},  M_{\xx} \right\rangle = 0$.
    \item Degeneracy with respect to energy $\left\langle \Xi_{\xx^*}, E_\xx \right\rangle = 0$.
    \item $\Xi[\xx^*]$ be even with respect to time-reversal transformation, see \cite{pavelka2014time}.
\end{enumerate}
The irreversible part of the evolution of a functional $F(\xx)$ is then given as
\begin{align}
    \left( \dot{F}(\xx) \right)_\mathit{irr}= \left\langle F_\xx,\ \Xi_{\xx^*}\middle|_{\xx^* = \frac{\delta S(\xx)}{\delta \xx}}\right\rangle~.
    \label{GDE01}
\end{align}
Gradient dynamics automatically satisfies the second law of thermodynamics (growth of entropy in isolated systems). This is guaranteed for instance for convex dissipation potentials, but also non-convexity far from the origin (equilibrium) be taken into account \cite{nonconvex}. Moreover, it is in close relation to the method of entropy production maximization \cite{Adam-EPM}. Gradient dynamics plays a key role when formulating dissipation in the GENERIC framework. 

Entropy-conjugate variables $\xx^*$ are most suitable for formulation of the gradient dynamics. However, it is often more straightforward to use the energy-conjugate variables $\xx^\dagger$, especially when combining the irreversible evolution with the Hamiltonian evolution (see e.g. \cite{be,ADER,ader-em,godr}). Details on the transformation between the conjugate variables can be found in \cite{pavelka2018multiscale,CMAT2018}.

A simple notorious example of the dissipation potential expressed in the energy-conjugate variables is 
\begin{align}     
    \Xi^\textrm{q} = \int\dd\rr \frac{1}{2\tau} 
    \left(\frac{\xi^\dagger}{s^\dagger} \right)^2\quad
    \text{and}\quad 
    \frac{\delta \Xi^\textrm{q}}{\delta \xi^\dagger}  = \frac{1}{\tau}\frac{\xi^\dagger}{(s^\dagger)^2} ~.     
\label{GDE02} 
\end{align} 
This is a prototype of dissipation potential, since any general dissipation potential can be approximated by a quadratic one due to the convexity near equilibrium and flatness at equilibrium. For further details see~\cite[Sect.\ 4.5]{pavelka2018multiscale}.
\subsection{Dynamic maximum entropy principle}\label{sec.dynmaxent}
The principle of maximum entropy (MaxEnt), where unknown value of a variable is determined by finding the maximum value of entropy subject to constraints given by declared knowledge, has been successfully applied in many fields (information theory, thermodynamics, etc.) \cite{jaynes}. However, in non-equilibrium thermodynamics the problem is not only to find value of a fast variable that has relaxed, but also to find the vector field along which the fast variable evolves, its evolution equation, when only less detailed variables are among the state variables (observables). To this end we recall the method of Dynamic MaxEnt (DynMaxEnt) \cite{grmela2013-cmat,pavelka2018multiscale}.
Extensive treatment of the DynMaxEnt principle in the context of the continuum thermodynamics can be found in~\cite{klika2019dynmaxent}.

\subsection{Relaxation of conjugate polarization momentum $\ppi$}
In~\autoref{sec.pol} polarization was equipped with its conjugate momentum $\ppi$. Inspired by the relaxation of the damped particle in \autoref{sec.dynmaxent}, we shall let the conjugate momentum relax to recover dissipative evolution of polarization. Let us choose state variables $(\rho_\alpha, \mm, s, \bcDD, \BB, \PP, \ppi)$, represented in~\eqref{TC11}. The energy depends on all the variables unless assumed otherwise.

\subsubsection{Polarization relaxation via $\ppi$}
Let us assume a dissipation potential quadratic in $\ppi^*$,
\begin{align}
	\Xi^{(\ppi)} = \int\dd\rr \frac{1}{2} \ppi^* \cdot \Lambda_{(\ppi)} \cdot \ppi^*
	= \int\dd\rr \frac{1}{2} \frac{\ppi^\dagger}{s^\dagger}\cdot \Lambda_{(\ppi)} \frac{\ppi^\dagger}{s^\dagger},
\end{align}
where $\Lambda_{(\ppi)}$ is a positive definite symmetric constant tensor.
Assuming also energy quadratic in $\ppi$, the MaxEnt value of $\ppi$ is zero. Using~\eqref{TC05}, the DynMaxEnt relaxation of $\ppi$ can be formulated as
\begin{align}\label{eq.mu.relax}
   0 = \partial_t \pi_i &= \cDD_i^\dagger - P_i^\dagger - \Lambda_{(\ppi) ij}\frac{\pi^{\dagger j}}{s^\dagger},
\end{align}
which is the constitutive relation to be plugged into the remaining evolution equations.
In particular, evolution equation for polarization \eqref{GG02P} becomes
\begin{align}
    \label{eq.P.relax}
    \partial_t P^i =  s^\dagger\Lambda_{(\ppi)}^{-1 ij}(\cDD_j^\dagger -  P_j^\dagger)-\partial_j\left(P^i v^j - P^j v^i \right) - v^i \partial_j P^j~,
\end{align}
which is an equation for polarization equipped with both reversible and irreversible terms.

Note that the irreversible terms on the right hand side of equation~\eqref{eq.P.relax} are generated by dissipation potential $\Xi^{(\ppi)}$ evaluated at constitutive relation \eqref{eq.mu.relax},
\begin{align}
	\Xi^{(\PP)} = \Xi^{(\ppi)}|_{\ppi^\dagger=\Lambda^{-1}_{(\ppi)} s^\dagger (\bcDD^\dagger-\PP^\dagger)} 
	=\int\dd\rr \frac{\Lambda^{-1}_{(\ppi)}}{2}\left(\PP^\dagger-\bcDD^\dagger\right)^2.%
    \label{PCR01}
\end{align}
Similar ideas were presented in~\cite{entrpo}.
This is the dissipation potential generating irreversible evolution of $\PP$ and $\bcDD$.
We may also pursue further relaxation of the polarization. If DynMaxEnt value of $\PP$ is zero, e.g. the total energy is quadratic in polarization, all the momentum coupling terms from~\eqref{eq.P.relax} vanish and we obtain
\begin{align}
    \bcDD^\dagger = \PP^\dagger
    \label{PCR07}
\end{align}
as the ultimate value of $\PP^\dagger$.

\subsubsection{Linear isotropic dielectric}\label{sec.dielectrics}
Let us now demonstrate the results of the previous paragraph on a particular choice of energy \eqref{eq.E.EMHDPmu} restricted to state variables $(\rho_\alpha,\vv,s,\bcDD,\BB,\PP)$.
Assuming that the internal energy be independent of $\PP$, the MaxEnt value of $\PP$ is zero so that~\eqref{PCR07} holds. We are going to determine the energy on the level where $\PP$ is relaxed.
Since the following considerations are simpler with velocity $\vv$ instead of momentum $\widehat\mm$, transformation~\eqref{eMIX17} is carried out.
Recalling~\eqref{PCR07}, we have
\begin{align}
    \frac{\PP}{\varepsilon_0\chi} = \PP^\wdagger =\PP^\dagger= \bcDD^\dagger 
= \bcDD^\wdagger+ \uu^\wdagger\times\BB \implies \PP = \chi \eepsilon_0 \left(\EE+\vv\times\BB\right)~.
    \label{PCR05}
\end{align}
Let us first observe that the relation~\eqref{PCR05}\textsubscript{right} can be also reformulated, using~\eqref{eMIX15b}, as
\begin{align}
    \PP = \chi \bcDD~,
    \label{PCR21}
\end{align}
Energy \eqref{eq.E.EMHDPmu} then becomes
\begin{align} 
    E_{\text{(relax)}} \rightarrow \int\textrm{d}\textbf{r}
    \left(\frac{1}{2}\rho \vv^2
	+\frac{1}{2}\bcDD\cdot\eepsilon^{-1}_0(1+\chi)\bcDD
	+\frac{1}{2}\BB\cdot\mmu^{-1}_0\BB
	+\vv\cdot(\bcDD\times\BB)
        +\eps(\rho,s)\right)~.
\end{align}
The relations~\eqref{TC01} and~\eqref{PCR21} can be combined to obtain,
\begin{align}
    \PP = \frac{\chi}{1+\chi}\DD~.
    \label{PCR06}
\end{align}
Finally, for either negligible magnetic field $\BB=0$ or negligible velocity $\vv=0$ we obtain that
\begin{align}
    \label{PCR22}
    \PP = \chi \eepsilon_0 \EE~,
\end{align}
which leads to the usual form of Coulomb's law for linear isotropic dielectrics~\cite{purcell1985electricity},
\begin{align}
\rho_\textrm{f} = \varepsilon_0\dive\big((1 + \chi) \EE\big)~,
\end{align}
since then $\bcDD^\dagger = \bcDD^\wdagger = \eepsilon_0\EE$.




\subsubsection{Single Relaxation Time model}
Consider again equation~\eqref{eq.P.relax} and assume further that $\vv=0$ and $\BB^\dagger$=0 and that the internal energy is negligible.
Then, for energy
\begin{align}
    E^\text{SRT} = \int\textrm{d}\textbf{r}\left(\frac{\bcDD^2}{2\epsilon_0}+\frac{\PP^2}{2\epsilon_0 \chi}\right)
  \label{SRAT04}
\end{align}
the evolution of polarization \eqref{eq.P.relax} becomes
\begin{align}
    \partial_t P^i &=  s^\dagger \Lambda^{-1}_{(\ppi)}\left(E_i - \frac{1}{\epsilon_0\chi} P_i\right),
    \label{SRAT01}
\end{align}
which represents a dissipative evolution of polarization subject to electric field.

Let us first analyze the evolution equation by applying harmonic electric field $\EE = {\EE}_0 \exp(i\omega t)$. Equation~\eqref{SRAT01} then gives the Single Relaxation Time (SRT) model of polarization, see e.g.~\cite{boettcher1979dielectric2},
\begin{align}
    \PP_0  = \frac{\chi \EE_0 }{1 + \varepsilon_0\frac{i \omega \chi}{\Lambda^{-1}_{\ppi} s^\dagger  }},
\label{SRAT02}
\end{align}
provided that $ \PP = \PP_0 \exp(i \omega t)$. 

In summary, by letting the conjugate polarization momentum $\ppi$ relax, a dissipative evolution of polarization is obtained \eqref{eq.P.relax}.
If the mechanical equilibrium is further assumed, this dissipative evolution is compatible with the SRT model widely used for comparison to experiments. 
Finally, the equilibrium of the dissipative evolution is the linear relation between polarization and electric intensity known from electrostatics.

\subsection{Relaxation of magnetization}
Let us now discuss relaxation of magnetization $\MM$ inspired by the Landau \& Lifshitz model \cite{LL1935}.
We consider dissipation potential
\begin{align}
    \Xi^\MM = \int\dd\rr \frac{1}{2} \left(\frac{\MM \times \MM^\dagger}{s^\dagger} \right)\cdot\LambdaM\cdot\left(\frac{\MM \times \MM^\dagger}{s^\dagger} \right),
    \label{DM01}
\end{align}
where $\LambdaM$ is a positive definite symmetric tensor field. Let us, for simplicity, assume that $\LambdaM$ is a constant multiple of the constant Eucledian metric and let us not distinguish between the upper and lower indexes in this section.
The derivative of~\eqref{DM01} w.r.t $\MM^\dagger$ is
\begin{align}
\frac{\delta\Xi^\MM}{\delta M_i^\dagger}%
=\frac{\LambdaM}{ {s^\dagger}^2}\varepsilon_{smi} M_m \varepsilon_{sjk} M_j M_k^\dagger%
= -\frac{\LambdaM}{ {s^\dagger}^2}\left( \MM \times(\MM\times\MM^\dagger) \right)_i~.
    \label{DM02}
\end{align}
Hence, the irreversible evolution of $\MM$ due to~\eqref{DM01} is 
\begin{eqnarray}
	\left( \partial_t M_i \right)_\mathit{irr} &=& -s^\dagger\frac{\delta\Xi^\MM}{\delta M_i^\dagger} =\frac{\LambdaM}{s^\dagger}\varepsilon_{ims} M_m \varepsilon_{sjk} M_j M_k^\dagger~\nonumber\\
	&=& \frac{\LambdaM}{s^\dagger} \left((\MM\cdot\MM^\dagger) \MM - (\MM\cdot\MM) \MM^\dagger\right),
    \label{DM03}
\end{eqnarray}
which is compatible with the Landau \& Lifshitz model of magnetization once suitable energy is provided, ~\cite[Sec.\,3.7]{brown1963micromagnetics}. 

Having recovered the Landau \& Lifshitz model, let us also formulate its generalized version advected by fluid mechanics and interacting with magnetic field.
Let us assume the following energy
\begin{align}
	\label{MB01a}    E^{(\MM)} = \int\dd\rr 
  \left(
  \frac{1}{2}\rho\vv^2 + \frac{\bcDD^2}{2\epsilon_0}  + \frac{(\BB - \mu_0 \MM)^2}{2\mu_0}+ \frac{\mu_0 \MM^2}{2\chi_\mathrm{m}}+ \vv\cdot(\bcDD\times\BB)\right)~.%
\end{align}
The energy-conjugate to the magnetization $\MM$ then reads
\begin{align}
    \label{MB01b}    \MM^\dagger  
=-\mu_0 \HH + \frac{\mu_0 \MM}{\chi_m}.
\end{align}
The field of \textit{magnetic intensity}, which is conjugate to $\BB$, satisfies 
\begin{align}
    \HH \stackrel{\mathrm{def}}{=} \BB^\wdagger = \frac{\BB}{\mu_0} - \MM,
\end{align}
which is the usual relation between $\HH$, $\BB$ and $\MM$ in the frame co-moving with matter. 
It holds, moreover, that
\begin{align}
    \dive\HH = -\dive\MM~.
    \label{MB02}
\end{align}
Combining~\eqref{eq.M.gen.v} and~\eqref{DM03}, we obtain the evolution of magnetization,
using also~\eqref{MB01b},
\begin{multline}
\partial_t M_i %
= -\gamma\varepsilon_{ijk} M_j B^k  - \gamma M_j\partial_i v^j - \gamma\partial_j\left( M_i v^j\right)\\%
-\frac{\LambdaM}{ s^\dagger}\varepsilon_{ims} M_m \varepsilon_{sjk} M_j  B^k ~.
    \label{MB03}
\end{multline}
Equation~\eqref{MB03}, supplied with the rest of Eqs. \eqref{GG02P}, is the generalized Landau-Lifshitz magnetization relaxation model\footnote{To fully recover the model, one should also add a weakly non-local term $\frac{1}{2}\alpha (\nabla\MM)^2$ to the energy density, see Sec. \ref{sec.mag}.}, where magnetization relaxes, interact with electromagnetic field and where it is advected by the fluid. 

\subsubsection{Linear isotropic magnetizable medium}
Let us demonstrate some properties of energy $E^{(\MM)}$ in the context of MaxEnt. Introducing the MaxEnt value of the magnetization,
\begin{align}
\label{MB10}    \frac{\delta E^{(\MM)}}{\delta \MM} = 0 \implies \MM^{\text{MaxEnt}} = \frac{\chi_m}{\mu_0(1 + \chi_m)}\BB~,
\end{align}
into the energy $E^{(\MM)}$ gives
\begin{align}
	\label{MB10a}    \widetilde E^{(\MM)} \to \int\dd\rr 
  \left(\frac{1}{2}\rho \vv^2 + \frac{\bcDD^2}{2\epsilon_0} + \frac{\BB^2}{2\mu_0(1+\chi_m)}+ \vv\cdot(\bcDD\times\BB\right)~,
\end{align}
for which it holds that
\begin{align}
    \label{MB11}
\BB = \mu_0(1+\chi_m)\HH~.
\end{align}
In particular, $\MM^\dagger = 0$ we obtain $\MM = \chi_m \HH$.
The MaxEnt value of the magnetization is used to recover the usual relation between the magnetic field and its intensity for linear isotropic materials.

\subsection{Electro-diffusion -- dissipation of $\bcDD$ and $\rho_\alpha$}\label{sec:electrodiff}
Let us now consider only state variables $(\rho_\alpha, \bcDD)$, which are essential for electrodiffusion. Poisson bracket~\eqref{GG01} for mixtures may be endowed with a weakly non-local electro-diffusion dissipation potential expressing friction between components of the mixture and the zero-th species (e.g. solvent),
\begin{align}
    \Xi^{\mathrm{D}}(\rho_\alpha^*, \bcDD^*) = \int\dd\rr\sum_{\alpha,\beta\neq 0}^n 
    &\left[\left( \partial_i\left(\rho_\alpha^* - \rho_0^*\right) - e_0\left(\frac{z_\alpha}{m_\alpha}-\frac{z_0}{m_0}\right) \cDD_i^*\right)\right.\nonumber\\
    &\left.\frac{M_{\alpha\beta}}{2}\left( \partial_i\left(\rho_\beta^* - \rho_0^*\right) - e_0\left(\frac{z_\beta}{m_\beta}-\frac{z_0}{m_0}\right) \cDD_i^*\right)\right],
    \label{EDD01}
\end{align}
where $M_{\alpha\beta}$ is a symmetric, positive definite matrix of binary diffusion coefficients. Note that $\nabla \rho^*_\alpha - e_0 \frac{z_\alpha}{m_\alpha}\cDD_i^*$ is proportional to the gradient of electrochemical potential of species $\alpha$.
This dissipation potential conserves both mass and energy.

The irreversible part of the evolution equations can be expressed in the energy-conjugate variables as 
\begin{subequations}\label{EDD.evo}
\begin{align}
    (\partial_t \rho_\alpha)_\textrm{irr} &= \sum_{\beta\neq 0}^n\partial_i\left( M_{\alpha\beta}\left( \partial_i \frac{\rho_\beta^\dagger - \rho_0^\dagger}{s^\dagger}%
    - e_0\left(\frac{z_\beta}{m_\beta}-\frac{z_0}{m_0}\right) \frac{\cDD_i^\dagger}{s^\dagger}\right)\right)\label{EDD04} \\
    (\partial_t \rho_0)_\textrm{irr} &= -\sum_{\alpha,\beta\neq 0}^n\partial_i\left( M_{\alpha\beta}\left( \partial_i \frac{\rho_\beta^\dagger - \rho_0^\dagger}{s^\dagger}%
    -e_0\left(\frac{z_\beta}{m_\beta}-\frac{z_0}{m_0}\right)  \frac{\cDD_i^\dagger}{s^\dagger}\right)\right)\label{EDD03} \\
    (\partial_t \cDD^i)_\textrm{irr} &=\sum_{\alpha,\beta\neq 0}^n e_0\left(\frac{z_\alpha}{m_\alpha}-\frac{z_0}{m_0}\right)%
    M_{\alpha\beta}\left( \partial_i\frac{\rho_\beta^\dagger - \rho_0^\dagger}{s^\dagger}%
    -e_0\left(\frac{z_\beta}{m_\beta}-\frac{z_0}{m_0}\right)  \frac{\cDD_i^\dagger}{s^\dagger}\right)\label{EDD02}~.%
\end{align}
\end{subequations}
Note that the divergence of~\eqref{EDD02} is equal to the sum of~\eqref{EDD03} and~\eqref{EDD04} weighted by the charge per mass of the respective species.
Therefore, the irreversible evolution given by~\eqref{EDD01} is compatible with Gauß's law given by~\eqref{GLEM07}.
In other words
\begin{align}
    \partial_i \left( (\partial_t \cDD^i)_\textrm{irr} \right) =  \sum_{\alpha=0}^n e_0\frac{z_\alpha}{m_\alpha}(\partial_t \rho_\alpha)_\textrm{irr}
    \label{EDD10}.
\end{align}
Equations \eqref{EDD.evo} express irreversible evolution of densities of the constituents and electric displacement under mutual interaction. 
The right hand sides of the equations for densities can be regarded as gradients of respective electrochemical potentials.
Evolution for the $\bcDD$ field leads to relaxation of the field consistent with the Gauß law and with the Poisson equation.

Note that a dissipation potential introducing dissipation of the partial mass densities (i.e.\ containing $\rho_\alpha^*$) is required to contain corresponding terms with $\bcDD^*$ otherwise the irreversible evolution would not be compatible with the Gauß law, c.f.~\eqref{EDD10}.%
Therefore, if the validity of Gauß's law for the free charge is required, then the form of dissipation potential involving $\bcDD^\dagger$ is constrained.

On short enough distances magnetic field effects are usually negligible compared with effects of the electric field, in contrast to long distances, where electric field usually does not play any relevant role due to screening. Let us analyze the former case, paving the way towards electrochemical problems.
The electro-diffusion due to~\eqref{EDD01} introduces dissipative fluxes identical to those of the generalized Planck-Nernst-Poisson systems (gPNP) presented in~\cite{dreyer2018bulksurface}.
Assuming that no magnetic field is present and that its evolution equation~\eqref{GLEM01} is satisfied, it follows that
\begin{align}
\DD^\dagger = -\nabla \varphi~,
\end{align}
see also~\eqref{PCR06}.
The considered level of description consists of $(\rho_\alpha, \uu, s, \varphi)$. 
Let us consider a further relaxation of the mass momentum $\uu$. We can proceed in two ways. Either  
we suppose for that energy does not depend on the mass momentum and the barycentric velocity vanishes,
or we can think of a dissipation of the mass momentum, e.g. viscosity, and use the DynMaxEnt principle in order to determine the value of the barycentric velocity $\uu^\dagger$. Let us now follow the latter route.
The energy is quadratic in $\uu$, see~\eqref{PCR06}, hence, its MaxEnt value is zero, leaving the remainder of the momentum equation as an constitutive equation for the velocity field. 
For the dissipation potential generating the irreversible part of the Navier-Stokes equations, see e.g.~\cite[Eqns.\ 4.74, 4.76]{pavelka2018multiscale}.
The viscous dissipative terms are then added to the reversible balance of the mass momentum. The whole gPNP-Stokes systems eventually reads
\begin{align}
    \partial_t \rho_\alpha%
    &= \partial_i\left(\rho_\alpha v_i + \sum_{\beta\neq 0}^n M_{\alpha\beta}\left( \partial_i \frac{\rho_\beta^\dagger - \rho_0^\dagger}{s^\dagger}%
    + e_0\left(\frac{z_\beta}{m_\beta}-\frac{z_0}{m_0}\right) \frac{\partial_i\varphi}{s^\dagger}\right)\right) \\
    \partial_t \rho_0%
    &= \partial_i\left(\rho_0 v_i -\sum_{\alpha,\beta\neq 0}^n M_{\alpha\beta}\left( \partial_i \frac{\rho_\beta^\dagger - \rho_0^\dagger}{s^\dagger}%
    +e_0\left(\frac{z_\beta}{m_\beta}-\frac{z_0}{m_0}\right)  \frac{\partial_i\varphi}{s^\dagger}\right)\right) \\
    0%
    & = -\sum_{\alpha=0}^n\rho_\alpha\partial_i\rho_\alpha^\dagger - s\partial_i s^\dagger - \rho_\mathrm{f} \partial_i \varphi
    +\partial_j\left( \frac{\widehat\mu}{2}\left(\partial_j v_i + \partial_i v_j \right) \right) + \lambda\partial_i\partial_j v_j\\
    0%
    &=-\partial_i \left( e_0(1+\chi)\partial_i \varphi \right)+ \sum_{\alpha=0}^n e_0\frac{z_\alpha}{m_\alpha}\rho_a.
\end{align}
Note that the third equation is not violated if the $curl$-operator is applied on it.


%
%
%
\section{Discussion}\label{sec.Dreyer}
W.~Dreyer, C.~Guhlke and R.~Müller in~\cite{dreyer2018bulksurface} published a comprehensive analysis of fluid mixtures coupled with electromagnetic fields, including polarization and magnetization which will be further referred as the DGM approach. Their treatment of surfaces, c.f. \cite{ondra-surface}, as independent thermodynamic systems interacting with the bulk, being beyond the scope of the presented work, have elucidated many electrochemical problems using non-equilibrium thermodynamics, for instance a unified theory of the Helmholtz and Stern layers, a derivation of Butler-Volmer equations, or useful asymptotic techniques, see~\cite{fuhrmann2016numerical,dreyer2016new,dreyer2018bulksurface,dreyer2014mixture,dreyer2014mixture,guhlke2015theorie}. 
Since our goal is in close relation to that works, let us compare the two approaches in detail. 

An important conceptual difference between the two approaches lies in the treatment of the state variables, i.e., the levels of the description. In our approach, the level of description is always defined first by declaring the set of state variables. Energy and entropy of the system can then, in principle, depend on all the state variables. 
Each state variable has its evolution given by the GENERIC equation~\eqref{eq.generic}, where the reversible and the irreversible parts of the evolution are separated. In contrast, the DGM approach develops a system of general balance equations which outlines the relations between the physical quantities and their fluxes and productions. The dependence of the fluxes and production upon the variables is determined using the entropy principle.

The DGM approach is formulated in an~\textit{inertial frame of reference} for which holds that (i) mass center not subjected to external forces moves with a constant velocity and (ii) the Lorentz-Maxwell-aether (or material) relations~\ref{eq.mat} are valid. This is similar, although not equivalent, to our construction. The choice of the energy in DGM,
\begin{align}
    \label{GRIL02}
    e^\text{D} = \frac{\uu^2}{2\rho}
    +\frac{\bcDD\cdot\EE}{2}
    +\frac{\BB\cdot\HH}{2}
                      +\epsilon^\text{D}%
\end{align}
is analogous to the energy~\ref{eq.E.EMHD}. Moreover, the formulation of the Maxwell equation in DGM is, in vacuum, formally equivalent to our approach.


In the non-vacuum case, the total charge is for both formulations given by $\dive\bcDD$. 
The $\dive\PP$ also bears the same meaning, i.e.~the density of the bounded charge.
The DGM evolution of the polarization $\PP$ is given as formal solution to the bound charge balance. It reads
\begin{align}
    \mbox{(DGM)}\quad    \partial_t P^i ={}& -\varepsilon_{ijk} \partial_j M_k^\text{D} + v_i \partial_j P^j + J_i^\textrm{P},
    \label{DPC01}
\end{align}
where $\MM^\text{D}$ is the~\textit{Lorentz magnetization} and $J^\textrm{P}$ denotes the non-convective part of the polarization current, cf.\ equation~\eqref{GG02P}.
The DGM formulation contains no balance equation concerning $\MM^\text{D}$, an evolution equation is later on found for $\widehat{\MM}^\text{D} = \MM^\text{D} + \vv\times\PP$ as a consequence of the closure. 
In contrast, the polarization bracket~\eqref{HPM01} contains the polarization momentum $\ppi$ as a variable and therefore implies its reversible evolution.
The polarization momentum $\ppi$ can be projected to a reduced variable $\boldsymbol{\mathcal{M}} = \curl \ppi$, would lead to the appearance of $\curl\boldsymbol{\mathcal{M}}^\dagger$ in the equation for polarization. This is perhaps the closest the two approaches can get in this respect. 

In DGM the balance equations for the densities of the partial masses $\rho_\alpha$, mass momentum $\uu$, total momentum $\uu+\bcDD\times\BB$, total energy $e^\text{D}$, internal energy $\epsilon^\text{D}$, total charge $\rho_\text{f} - \dive\PP$, bound charge $-\dive\PP$, flux of the magnetic field, and the entropy $s^\text{D}$ are formulated, see~\cite{dreyer2018bulksurface}, which is translated in our approach to the respective evolution equations. 
In particular, the total energy and the total momentum $\uu + \bcDD\times\BB$ are, in the absence of external forces, conserved quantities in DGM as in our approach. Coupling between the charged fluid and electromagnetic field is given by the choice of Lorentz's force~\cite[Eqn. 36a]{dreyer2018bulksurface}:
\begin{align} 
k_i = %
      \partial_j \cDD^j E_i + \varepsilon_{ijk} \left(v_j \partial_l \cDD_l + J^\textrm{F}_j + J^\textrm{P}_j\right)B^k 
    \label{DPC02}
\end{align}
as a source term in the mass momentum balance and by the choice of Joule heating
\begin{align}
    \pi = (v_i\partial_j \cDD^j  + J_i^\textrm{F} + J_i^\textrm{P}) E_i
    \label{DPC03}
\end{align}
as a source term in the internal energy balance, cf.\ Appendix~\ref{app:momentum}. Symbols $J^\rF$ denotes the non-convective part of the free charge current. This is similar to our formula \eqref{TC15}, at least in the absence of irreversible terms.

As it is shown in Appendix~\ref{app:momentum}, the presented coupling between the motion of mass and the electromagnetic field can be formulated using a similar choice of Joule's heating, see~\eqref{AMO18}, and restricts the form of the total momentum to~\eqref{AMO15}. The dissipation potentials for the irreversible, conductive, currents are formulated in terms of total momentum $\widehat\mm$, so that $\bcDD^\dagger$ is the co-moving electric field. Hence, the Lorentz force acting upon the conductive currents can be found when the dissipation potentials are transformed into the variables with the mass momentum $\uu$.

The choice of Lorentz's force~\eqref{DPC02} tells that the electric field acts upon the total charge and the magnetic field upon convective current density, $\mathbf{J}^\rF$ and $\mathbf{J}^\text{P}$ within the DGM theory. This is equivalent for the total charge and the conductive free charge $\mathbf{J}^\rF$. The difference appears for $\mathbf{J}^\text{P}$, since it is not purely irreversible and contains reversible terms in the sense of the time-reversal transform. 

Our evolution equation for polarization consists of reversible and irreversible parts. The former is given by Eq. \eqref{eq.P.gen.v}, and the terms of the right hand side containing velocity be rewritten as the Lie derivative of $\PP$, $\Lie_\vv \PP$. Polarization is thus simply advected by the fluid (apart from interacting with the electromagnetic fields and apart from relaxation processes). On the other hand, polarization is not simply advected in DGM as the equation for $\PP$ contains other velocity-dependent terms therein. This is a difference between our approach and DGM.

Authors of DGM assumed that the entropy density $s^\text{D}$ depends only on a specific subset of the variables 
\begin{align}
    s^\text{D}(\epsilon^\textrm{D} + \widehat\MM^\text{D}\cdot\BB, \rho_\alpha, \PP, \widehat\MM^\text{D}) \implies \frac{1}{T^\text{D}}=\frac{\partial s^\text{D}}{\partial(\epsilon^\text{D} + \widehat\MM^\text{D}\cdot\BB)}~, 
    \label{GRIL03}
\end{align}
thus obtaining a specific definition of the temperature $T^\text{D}$, see~\cite[Eqn. 48a, 49a]{dreyer2018bulksurface}\footnote{the dependence on the bulk deformation gradient is omitted}. 
 This choice allowed them to find a reasonably simple closure of the equations using the entropy principle, c.f. \cite{jou2010extended}. On the other hand, temperature is in our approach defined as derivative of the internal energy density with respect to entropy as is usual \cite{degroot1984nonequilibrium}.

Let us now focus on some features of the closure, especially, the non-convective flux of the bound charge that.
They derived the following evolution of the polarization $\PP$ and magnetization $\widehat\MM^\text{D}$,
\begin{subequations}
\begin{align}
\partial_t P^i =%
    {}& \underbrace{
        -v_j\partial_j P^i 
        + \frac{1}{2}P^j\left(
              \partial_j v_i - \partial_i v_j
          \right) 
        + \frac{1}{\tau^\textrm{P}}\left(
              T^\text{D}\frac{\partial s^\text{D}}{\partial P^i}+ E_i + \varepsilon_{ijk}v_j B^k 
          \right)
      }_{={J}^\textrm{P}_i - \varepsilon_{ijk}\partial_j M_k^\text{D} + {v}_i\partial_j P^j}~,\label{DPC23}\\ 
	\label{eq.DGM.M}
\partial_t \widehat M_i^\text{D} =%
    {}&-v_j\partial_j \widehat M_i^\text{D} 
        + \frac{1}{2}\widehat M_j^\text{D}\left(
              \partial_j v_i - \partial_i v_j
          \right) 
        + \frac{1}{\tau^\textrm{M}}
              \left(
                    T^\text{D}\frac{\partial s^\text{D}}{\widehat M_i^\text{D}} + B^i \right)~,
\end{align}
respectively. The phenomenological coefficients $\tau^\textrm{P}$ and $\tau^\textrm{M}$ are relaxation times of polarization and magnetization, respectively. 
\end{subequations}
The relaxation of the polarization, i.e.~the last bracket in~\eqref{DPC23}, is equivalent to the dissipation derived using the DynMaxEnt principle given by potential $\Xi^{\PP}$, see~\eqref{PCR01}.
The conductive flux of free charge $\mathbf{J}^\textrm{F}$ is equivalent to the one generated by the dissipation potential $\Xi^\textrm{D}$ in~\eqref{EDD02}, which also holds for the mass diffusion fluxes.

When the relaxation time of polarization $\tau^\textrm{P}$ can be considered large, the dissipative part in~\eqref{DPC23} can be neglected. Eventually, the reversible part of the bound charge evolution in a volume $V$ for the two theories read
\begin{align}
    \label{DPC08}
\mbox{(DGM)}\quad    \partial_t \int_V\partial_i P^i 
    &= \int_{\partial V} \left(
        -v_j\partial_j P^i 
        + \frac{1}{2}P^j\left(\partial_j v_i 
        - \partial_i v_j\right) 
    \right)\nu_i~,\\
	\mbox{(our)}\quad    \partial_t \int_V \partial_i P^i
    &= \int_{\partial V} \left( 
        \pi^{\dagger i} 
        - v^i\partial_j P^j
    \right)\nu_i~.
\end{align}
The net bound charge in a fixed volume $V$ can thus be reversibly changed by convection or by change of $\ppi^\dagger$. In contrast to DGM, the bound charge is advected like a scalar field in our approach, which corresponds with that polarization is Lie-dragged in our approach.


Also the evolution of magnetization in DGM, Eq. \eqref{eq.DGM.M}, is different from our equation \eqref{eq.M.gen.v}. In our case the evolution of magnetization caused by fluid motion is again just the Lie drag. Apart from that, there is a contribution from inertia of the magnetization itself, as in the Landau-Lifshitz model, which brings interaction with the magnetic field when energy depends on both magnetization and magnetic field. 

Let us consider the volume and shear viscosity to be vanishing, the stress tensor presented in~\cite[Eqn 62]{dreyer2018bulksurface} reads\footnote{omitting also the dependence on the volume-preserving deformation gradient}
\begin{align}
    \mbox{(DGM)}\quad\sigma_{ij} %
    &= p(T^\text{D}, \rho_\alpha, \widehat{\MM}^\text{D}, \PP)\delta_{ij} 
+\left(\widehat{M}^\text{D}_k B_k +  \widehat{E}_k P_k   \right)\delta^{ij}\nonumber \\
&+\frac{1}{2}\left(\widehat{E}_i P_j +  P_i\widehat{E}_j   \right)
-\frac{1}{2}\left(\widehat{M}^\text{D}_i B_j +  B_i\widehat{M}^\text{D}_j   \right)
    \label{}
\end{align}
here $\widehat{\EE} = \EE + \vv \times\BB$. The dependence of the stress tensor $\sigma$ on the material properties linked to the polarization $\PP$ and magnetization $\widehat\MM^\text{D}$ is provided by the isotropic part $p\delta_{ij}$ the rest of the tensor is linear in $\PP$ and $\widehat{\MM}^\text{D}$, independent of the choice of the energy. An energy weakly non-local in polarization or magnetization, see e.g.~\eqref{PW01} and~\eqref{MSI01}, leads in the here presented treatment to a structurally different stress since a non-symmetric components can appear in the evolution equation for the mass momentum $\uu$, see~\eqref{PW04} or~\eqref{MSI04}. The stress in DGM is different from our formulas for stress, mainly in the off-diagonal part.

The Onsager-Casimir reciprocal relations \cite{Onsager1931,casimir1945onsager,degroot1984nonequilibrium} are one of the corner-stones of non-equilibrium thermodynamics.
Roughly speaking, they say that
state variables with the same parity are coupled through an operator symmetric with respect to the simultaneous transposition and time-reversal while variables with opposite parities are coupled
through an antisymmetric operator. 
 They are automatically satisfied within the GENERIC framework in a generalized sense (beyond near equilibrium) \cite{ottinger2005minimal,pavelka2018multiscale,pavelka2014time}. In \cite{dreyer2018bulksurface} they seem to be satisfied as well, but parities of the state variables are determined from the power of seconds in the units of the variables instead of the time-reversal transformation, which is not invariant with respect to changing the physical units. Although the results seem to be all right, one should be careful in principle, perhaps using a precise definition of affine and vector spaces \cite{Matolcsi,tamas-kinematics} or the definition based on projections from more detailed levels \cite{pavelka2018multiscale}.

\section{Summary and Conclusion}
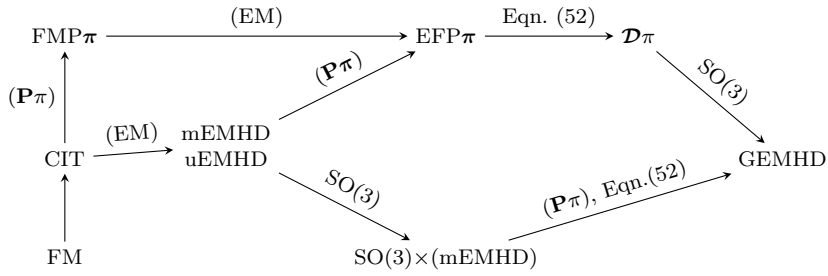
\begin{figure}
\begin{tikzpicture}
  \matrix (m) [matrix of nodes, row sep=3em, column sep=3em]{
    & \node[] (FMPPpi){FMP$\ppi$};
    &
    & \node[] (EFPPpi){EFP$\ppi$}; 
    & \node[] (bcDDpi){$\bcDD\pi$};
    \\

    & \node[] (CIT) {CIT}; 
    & \node[align=center] (EMHD) {mEMHD\\uEMHD}; 
    & 
    & 
    & \node[] (GEMHD){GEMHD};\\
    & \node[] (FM){FM};
    & 
    & \node[] (SO3MD)  {SO(3)$\times$(mEMHD)}; \\
    };
   \path[-stealth] (FM)     edge (CIT) ;
   \path[-stealth] (CIT)    edge node[midway,above,sloped] {(EM)} (EMHD);
   \path[-stealth] (CIT)    edge node[midway,left] {($\PP\pi$)} (FMPPpi);
   \path[-stealth] (FMPPpi) edge node[midway,above] {(EM)} (EFPPpi);
   \path[-stealth] (EMHD)   edge node[midway,above,sloped] {($\PP\ppi$)} (EFPPpi);
   \path[-stealth] (EFPPpi) edge node[midway,above,sloped] {Eqn. \eqref{TC01}} (bcDDpi);
   \path[-stealth] (bcDDpi) edge node[midway,above,sloped] {SO(3)} (GEMHD);
   \path[-stealth] (EMHD)   edge node[midway,above,sloped] {SO(3)} (SO3MD);
   \path[-stealth] (SO3MD)  edge node[midway,above,sloped] {($\PP\pi$), Eqn.\eqref{TC01}}(GEMHD);
\end{tikzpicture}
\caption{\label{fig:hierarchy} Hierarchy of the Poisson brackets. The Poisson bracket CIT is gradually extended using the semidirect product eventually reaching the over-arching bracket GEMHD.}
\end{figure}

In the first section a hierarchy of Poisson brackets describing the reversible dynamics of a charged, polarized and magnetized continua coupled with electromagnetic field has been developed by means of differential geometry, see Figure~\ref{fig:hierarchy}.
The semidirect product of the fluid mechanics Lie Algebra dual $(\rho, \uu, s)$ and the electromagnetism cotangent bundle $(\AA,-\DD)$ results in the reversible electro-magneto-hydrodynamics already known~\cite{marsden1982hamiltonian,holm1986hamiltonian,esen2017hamiltonian,pavelka2018multiscale}.
Newly, cotangent bundle $(\ppi,-\PP)$ describing dynamics of bound charge is also coupled to $(\rho, \uu, s, \DD,\BB)$ using the same technique.
Finally, the classical spin dynamics represented by the local Lie algebra dual of $\text{SO}(3)$ is coupled to the Lie algebra dual of fluid mechanics, which can be seen as advection of the Landau-Lifshitz model by fluid mechanics. The theory is the Galilean invariant.

The second section is dedicated to introduction of irreversible dynamics using dissipation potentials and subsequent reductions of the before-built levels of description to the less detailed levels. The reduction is carried out by the Dynamic Maximization of Entropy (DynMaxEnt) technique, which forms passage from finer to rougher levels of description in a geometric way.
In particular, the polarization momentum $\ppi$ is relaxed giving rise to dissipation of the polarization field itself.
Further exploitation of the induced dissipation of $\PP$ leads to the standard formulas for linear dielectrics and to the Single Relaxation Time model. 
The dissipation potential for the magnetization $\MM$ is found so that the Landau-Lifshitz model of spin relaxation is restored.
Finally, the electro-diffusion dissipation potential is introduced, leading to a generalized Poisson-Nernst-Planck-Stokes model.

In this rather complex dynamics, we observe phenomena like advection of polarization and magnetization by the fluid, polarization and magnetization waves and the respective inertial effects. We leave concrete applications of these phenomena for future researchers.

\section{Acknowledgment}
Authors thank to professor František Maršík for his relentless encouragement to explore the relation of electromagnetism and non-equilibrium mixture theories on a fundamental level.
Authors thank also to Clemens Guhlke, Rüdiger Müller and Wolfgang Dreyer for their utter patience and openness during the elucidating discussions on the topic.
We are also grateful to Péter Ván and Tamás Fülöp for consulting response to a reviewer regarding electrodynamics and space-time.
PV and MP are indebted to Miroslav Grmela for teaching us to geometric thermodynamics.

This work was supported by the German Research Foundation, DFG project no.~FU~316\/14-1.
This work was supported by the Czech Science Foundation, project no.~17-15498Y.
This work was supported by Charles University Research program No. UNCE/SCI/023.



\appendix

\section{Notation}

\begin{table}[h]
\caption{List of the basic quantities, symbols and SI units.}
	 \begin{tabular}{@{}lll@{}}
  Quantity & Symbol & SI units\\
\hline\noalign{\smallskip}
  {magnetic field}          & $\BB$               & \SI{}{\kilogram\per\second^2\per\ampere} \\
  {magnetic field intensity}& $\BB^\dagger,\HH$   & \SI{}{\ampere\per\meter} \\
  {displacement field}      & $\DD$               & \SI{}{\coulomb\per\meter^2} \\
  {electric induction field}& $\bcDD$              & \SI{}{\coulomb\per\meter^2} \\
  {electric field}          & $\EE, \DD^\dagger, \bcDD^\dagger$               & \SI{}{\volt\per\meter} \\
  {electric potential}      & $\varphi$           & \SI{}{\volt} \\
  {polarization density}    & $\PP$               & \SI{}{\coulomb\per\meter^2} \\
  {polarization momentum }  & $\ppi$              & \SI{}{\volt\second\per\meter} \\
  {magnetization density}   & $\MM$               & \SI{}{\ampere\per\meter} \\
  {partial mass density}		& $\rho_\alpha$	      & \SI{}{\kilogram\per\meter^3}\\
  {chemical potential}		  & $\rho_\alpha^\dagger$& \SI{}{\joule\per\kilogram}\\
  {mass momentum density}		& $\uu$	   	          & \SI{}{\kilogram\per\meter^2\per\second}\\
  {total momentum density}  & $\mm, \widehat\mm$	& \SI{}{\kilogram\per\meter^2\per\second}\\
  {entropy density}         & $s$                 & \SI{}{\joule\per\kelvin\per\meter^3}\\
 \noalign{\smallskip}\hline
\end{tabular}
\label{notation}
\end{table}

\begin{table}[h]
\caption{List of momenta densities and the corresponding levels of description.}
	 \begin{tabular}{@{}rl@{}}
       momentum density & level of description\\
\hline\noalign{\smallskip}
  $\uu$	   	      & $(\rho_\alpha, s, \uu)$\\
  $\mm$	   	      & $(\rho_\alpha, s, \mm, \DD, \BB)$\\
  $\widehat\mm$	  & $(\rho_\alpha, s, \widehat\mm, \PP, \ppi, \bcDD, \BB, \MM)$\\
 \noalign{\smallskip}\hline
\end{tabular}
\label{mnotation}
\end{table}

\section{Elementary dipole}
Two classical charged mass points are described by their positions, $\rr^1$ and $\rr^2$, momenta $\pp^1$ and $\pp^2$, carrying charge $q_1$ and $q_2$, respectively.
The dynamics of the particles is governed by the canonical Poisson bracket:
\begin{align}
    \{F,G\}^\textrm{mp}(\rr^\beta,\pp^\beta) = \sum_{\beta=1}^2 (F_{r_i^\beta} G_{p_i^\beta} - G_{r_i^\beta} F_{p_i^\beta})~.
  \label{EDI01}
\end{align}
Assume that $q = q_1 = -q_2$ and consider the following transformation of the variables:
\begin{align}
    \RR = \frac{m_1\rr^1 + m_2\rr^2}{m_1+m_2},\quad \mathcal{P} = q(\rr^1 - \rr^2)\label{EDI04},\\
    \Pi = \pp^1 + \pp^2,\quad \textbf{t} = \frac{m_2\pp^1 - m_1\pp^2}{q(m_1 + m_2)}\label{EDI08},
\end{align}

Bracket~\eqref{EDI01} then transforms into:
\begin{align}
    \{F,G\}^\textrm{mp}(\RR,\Pi,\mathcal{P},\mathbf{t}) =& (F_{R_i} G_{\Pi_i} - G_{R_i} F_{\Pi_i})\nonumber\\
    &\phantom{F_{\mathcal{P}_i} G_{t_i}}+(F_{\mathcal{P}_i} G_{t_i} - G_{\mathcal{P}_i} F_{t_i})~.
  \label{EDI02}
\end{align}

\section{Galilean transformations}\label{sec.Gal}

Maxwell equations \eqref{EM05} are not in the usual form, e.g. from \cite{Landau2}. Usually the equations are derived in the special relativistic setting by first forming the electromagnetic four-cotensor from fields $\EE$ and $\BB$, and the resulting evolution equations are for $\EE$ and $\BB$ as well. 

The Maxwell equations are naturally Lorentz invariant, as follows from the special relativistic treatment. 
However, it is often the case that Galilean invariance is sufficient for describing the observed phenomena. Moreover, Galilean transformations can be seen as low-velocity and low-strength limits of Lorentz transformations. Since Maxwell equations adequately describe usual phenomena around us, what is their behavior with respect to the Galilean transformations? 

When having the pair $(\EE,\BB)$ at hand, as in the usual case, one has to be restricted either to the electric limit (magnetic effects negligible) or magnetic limit (electric effects negligible), as discussed in \cite{leblond1973galilean}. However, when we have all the fields $(\EE,\BB,\DD,\HH)$, which is the case of Eqs.~\eqref{eq.gen.v}, the equations are invariant with respect to Galilean transformations
\begin{subequations}\label{eq.trafo.Gal}
	\begin{align} 
        \DD' = \DD, &\qquad \EE' = \EE + \VV\times \BB,\\
        \BB' = \BB, &\qquad \HH' = \HH - \VV\times \DD,\\
        \PP' = \PP, &\qquad \ppi' = \ppi,\\
        \MM' = \MM,
	\end{align}
which corresponds to the change of position $\rr=\rr' + \VV t$ and velocity $\vv = \vv'+\VV$. Note that polarization and magnetization are not affected by the transformation.
\end{subequations}
The reasoning can be found for instance in \cite{Matolcsi} or \cite{leblond1973galilean}, where four-vectors and four-covectors are split into the time-like and space-like parts, and the Galilean-invariant Maxwell equations then result from such splitting of the special relativistic four-dimensional structure in the low velocity limit, c.f. \cite{Matolcsi-Van}.

Maxwell equations in vacuum \eqref{EM05} are not Galilean invariant. However, equations \eqref{eq.EMHD.v}, where Maxwell equations are complemented by motion of matter, are Galilean invariant, as noted in \cite{leblond1973galilean}, albeit without the evolution equations for matter. In \cite{leblond1973galilean} they found material relations \eqref{eq.mat} not being Galilean invariant, since they had to be formulated in a special reference frame (e.g. aether). However, we have a natural reference frame at hand (the frame co-moving with the center of mass) where the material relations are valid. This makes the theory fully Galilean invariant.

\section{Relation of the mass momentum and the total momentum}\label{app:momentum}
Let us define the mass momentum as
\begin{align}
    \uu = \rho \mm^\dagger = \rho\mathbf{v},
    \label{AMO07}
\end{align}
for some given energy $E(\rho, \mm, s, \bcDD, \BB, \PP, \mmu, \MM)$.
This relation thus defines a function $\widetilde{\uu}$ such that 
\begin{align}
    \uu = \widetilde{\uu}(\rho, \mm, s, \bcDD, \BB, \PP, \mmu, \MM)~.
\end{align}
Let us assume that there exists a different representation of energy $\widetilde{E}(\rho, \uu, s, \bcDD, \BB, \PP, \mmu, \MM)$. 
However, the energy of the system is invariant w.r.t. the choice of representation. Therefore,
\begin{align}
    \widetilde{E}(\rho, \uu, s, \bcDD, \BB, \PP,\mmu,\MM) = E(\rho, \mm, s, \bcDD, \BB, \PP,\mmu,\MM)~.
    \label{AMO08}
\end{align}
We require the velocity field to be also invariant of the chosen representation, 
\begin{align}
    \mathbf{v} = \mm^\dagger = \uu^\wdagger~.
    \label{AMO09}
\end{align}
As a consequence of the chain rule and formula~\eqref{AMO07}, we obtain more specific form of $\widetilde\uu$, that is,
\begin{align}
    \widetilde{u}_i(\bcDD, \BB, \mm) = m_i + f_i(\bcDD, \BB)~,
    \label{AMO10}
\end{align}
the extra momentum $\mathbf{f}$ is also assumed to be independent of mass and entropy density and being caused only due to the electromagnetic field.

The form of~\eqref{AMO10} suggest that the extra momentum should be the momentum of the electromagnetic field.
Since the theory of the Maxwell equations is built upon the assumption of the field superposition, we assume $\mathbf{f}$ to be a bilinear function of $\bcDD$ and $\BB$, hence we have,
\begin{align}
    f_i = a_{ijk} \cDD^j B^k \quad\text{where}\quad a_{ijk} \in \mathbb{R}~.
    \label{AMO11}
\end{align} 

Using the chain rule, formulas~\eqref{AMO08} and \eqref{AMO11}, we find the relations between the derivatives of the energy in the two representations,
\begin{subequations}
    \begin{align}
        \rho^\dagger = \rho^\wdagger,\ s^\dagger = s^\wdagger,~\mm^\dagger = \uu^\wdagger~,\\
        \PP^{\dagger} = \PP^\wdagger,\ \mmu^{\dagger} = \mmu^\wdagger,~\MM^{\dagger} =  \MM^\wdagger~,\\
        \cDD_i^{\dagger} = \cDD_i^\wdagger - u_l^\wdagger a_{lik} B^k~,  \\
        B_i^\dagger = B_i^\wdagger - u_l^\wdagger a_{lji} \cDD^j~.
    \end{align}
    \label{AMO02}
\end{subequations}

The kinetic energy of the system may be defined using either the velocity or the mass momentum.
In general, kinetic energy is not a conserved quantity and may be changed during evolution of the system.
Experimental evidence suggests that the change of the kinetic energy of a charged matter, due to the presence of the electromagnetic field, is linearly proportional to the charge density and the electric field, i.e., $\partial_i \cDD^i \cDD^\dagger_j m^{\dagger j}$.
We require this power density to be independent of the chosen representation, i.e., $\partial_i \cDD^i \cDD^\dagger_j m^{\dagger j}= \partial_i \cDD^i \cDD^\wdagger_j v^j $.
Using~\eqref{AMO02} we have
\begin{align}
    \partial_i \cDD^i \cDD_j^{\dagger} {m^\dagger_j} =  \partial_i \cDD^i \left(\cDD_j^{\wdagger} - u^{l\wdagger}a^{lj}{}_{k} B^k\right) \widetilde u^{j\wdagger}~. 
    \label{AMO18}
\end{align}
To this end, the extra term on the right-hand side must be zero, i.e., it must vanish for arbitrary magnetic and velocity fields. For example, it must hold
\begin{align}
    0 = a_{ijk} v_i v_j\quad\forall k \in\{1,2,3\}~.
    \label{}
\end{align}
Apparently, the tensor $a_{ijk}$ is antisymmetric and isotropic. Up to a sign, there exists only one such tensor in three dimensions,
\begin{align}
    a_{ijk} = \varepsilon_{ijk}~.
    \label{AMO14}
\end{align}
The sign choice is due to the experimentally observed difference in bending of rays of negatively and positively charged particles~\cite{LL1935}. 
Let us the calculate time change of the kinetic energy,
\begin{subequations}
\begin{align}
    \partial_t \int \frac{\uu^2}{2\rho} &= \int \frac{\uu}{\rho}\left(\partial_t (\mm - \bcDD\times\BB) - \frac{\uu}{2\rho}\partial_t \rho\right)\nonumber\\
&=\int\dd\rr u_i^\wdagger
    \left(
      -\rho\partial_i \rho^\wdagger - s\partial_i s^\wdagger- u_j\partial_i {u_j}^\wdagger
        + \frac{1}{2} u_i^\wdagger\partial_j
                      \left(\rho u_j^\wdagger \right)
    \right)\label{AMO12}\\
&+\int\dd\rr u_i^\wdagger\left(
       -P^j\partial_i P_j^\dagger +\mu^{\dagger j}\partial_i\mu_j
  + \partial_j\left(
      P_i^\wdagger P^j -  \mu_i\mu_j^\wdagger
    \right)\right)\label{AMO17} \\
&+\int\dd\rr u_i^\wdagger\left(
            - M_j\partial_i M_j^\wdagger 
  - \partial_j\left(
      M_i M_j^\wdagger
    \right)
\right)\label{AMO16} \\
&+\int\dd\rr u_i^\wdagger\left(
      \cDD_i^\wdagger \partial_j \cDD^j 
    + B_i^\wdagger \underbrace{\partial_j B^j}_{=0} \right)\label{AMO13}~.
\end{align}
\end{subequations}
The terms on lines~\eqref{AMO12}, \eqref{AMO17}, \eqref{AMO16} are the changes of the kinetic energy due to the fluid mechanics, dipole mechanics and spin mechanics, respectively.

The terms on line~\eqref{AMO13} are the change of the kinetic energy due to the power of the electromagnetic field.
In total, the change of kinetic energy is independent of the chosen energy representation, see~\eqref{AMO02} with the choice of~\eqref{AMO14}.

In summary, the relation between the mass momentum $\uu$ and the total momentum $\mm$
\begin{align}
    \mm = \uu + \bcDD\times\BB~
    \label{AMO15}
\end{align}
is recovered.

\end{document}